\documentclass[trackchanges,twocolumn]{aastex701}

\usepackage[T1]{fontenc}
\usepackage{xspace}
\usepackage{amsmath}

\newcommand{\hoceans}{$H_{\mathrm{oceans}}$\xspace}
\newcommand{\water}{H$_{2}$O\xspace}
\newcommand{\cotwo}{CO$_{2}$\xspace}
\newcommand{\fotwo}{$f\mathrm{O}_{2}$\xspace}
\definecolor{mypink}{RGB}{255,105,180} 

\begin{document}
\title{Thermal Evolution of Lava Planets Across System Ages: \\
Predictions for \emph{Hell of a Survey}}

\author[orcid=0009-0008-7799-7976]{Mariana Sastre} 
\affiliation{Kapteyn Astronomical Institute, University of Groningen, Landleven 12, 9747 AD Groningen, The Netherlands}
\email[show]{m.c.villamil.sastre@rug.nl}  

\author[orcid=0000-0002-3286-7683]{Tim Lichtenberg} 
\affiliation{Kapteyn Astronomical Institute, University of Groningen, Landleven 12, 9747 AD Groningen, The Netherlands}
\email{tim.lichtenberg@rug.nl}

\author[orcid=0000-0003-4987-6591]{Lisa Dang}
\affiliation{Waterloo Centre for Astrophysics and Department of Physics and Astronomy, University of Waterloo, Waterloo, Ontario, Canada}
\email{lisa.dang@uwaterloo.ca}

\author[orcid=0000-0002-4487-5533]{Anjali A. A. Piette}
\affiliation{School of Physics and Astronomy, University of Birmingham, Birmingham, UK}
\email{a.a.a.piette@bham.ac.uk}

\author[orcid=0000-0001-8618-3343]{Haiyang S. Wang}
\affiliation{Center for Star and Planet Formation, Globe Institute, University of Copenhagen, Copenhagen, Denmark}
\email{haiyang.wang@sund.ku.dk}

\author[orcid=0000-0003-3204-8183]{Mercedes López-Morales}
\affiliation{Space Telescope Science Institute, 3700 San Martin Drive, Baltimore, MD 21218, USA}
\email{mlopez-morales@stsci.edu}

\author[orcid=0000-0001-8749-1962]{Thomas Wilson}
\affiliation{Department of Physics, University of Warwick, Gibbet Hill Road, Coventry CV4 7AL, UK}
\email{thomas.g.wilson@warwick.ac.uk}

\author[orcid=0000-0002-9265-4209]{Charles-Edouard Boukaré}
\affiliation{Department of Physics and Astronomy, York University, Toronto, Ontario, Canada}
\email{boukare@yorku.ca}

\author[orcid=0009-0004-3980-8143]{Mahesh Herath}
\affiliation{Trottier Space Institute, McGill University, 3550 Rue University, Montreal, QC, H3A 2A7, Canada}
\affiliation{Department of Earth \& Planetary Sciences, McGill University, 3450 rue University, Montr\'eal, QC H3A 0E8, Canada}
\email{mahesh.herath@mail.mcgill.ca}

\author[orcid=0000-0001-6129-5699]{Nicolas Cowan}
\affiliation{Department of Earth \& Planetary Sciences, McGill University, 3450 rue University, Montr\'eal, QC H3A 0E8, Canada}
\affiliation{Department of Physics, McGill University, 3600 rue University, Montr\'eal, QC H3A 2T8, Canada}
\email{nicolas.cowan@mcgill.ca}

\author[orcid=0000-0002-3696-2127]{Md Abdullah Al Zaman}
\affiliation{Department of Physics and Astronomy, University of Kansas, Lawrence, KS, USA}
\email{abdullah.alzaman@ku.edu}

\author[orcid=0000-0002-8399-472X]{Madyson G. Barber}
\affiliation{Department of Physics and Astronomy, The University of North Carolina at Chapel Hill, Chapel Hill, NC 27599, USA}
\email{madysonb@live.unc.edu}

\author[orcid=0000-0002-4480-310X]{Casey Brinkman-Traverse}
\affiliation{Trottier Space Institute, McGill University, 3550 Rue University, Montreal, QC, H3A 2A7, Canada}
\email{casey.brinkman@mcgill.ca}

\author[orcid=0000-0001-5848-6750]{Nicholas Connors}
\affiliation{Waterloo Centre for Astrophysics and Department of Physics and Astronomy, University of Waterloo, Waterloo, Ontario, Canada}
\affiliation{Department of Physics and Trottier Institute for Research on Exoplanets, Universit\'e de Montr\'eal, Montreal, QC, Canada}
\email{nicholas.connors@umontreal.ca}

\author{Ian Crossfield}
\affiliation{Department of Physics and Astronomy, University of Kansas, Lawrence, KS, USA}
\email{ianc@ku.edu}

\author[orcid=0009-0005-9175-4768]{Lina D'Aoust}
\affiliation{Waterloo Centre for Astrophysics and Department of Physics and Astronomy, University of Waterloo, Waterloo, Ontario, Canada}
\email{ldaoust@uwaterloo.ca}

\author[orcid=0000-0002-1807-4441]{Oliver Herbort}
\affiliation{Department of Astrophysics, University of Vienna, T\"urkenschanzstr. 17, 1180 Vienna, Austria}
\email{oliver.herbort@univie.ac.at}

\author[orcid=0009-0002-9902-731X]{Leoni Janssen}
\affiliation{Leiden Observatory, Leiden University, Leiden, The Netherlands}
\email{ljanssen@strw.leidenuniv.nl}

\author[orcid=0000-0003-1196-6233]{Mathilde Kervazo}
\affiliation{Laboratoire de Plan\'etologie et G\'eosciences, UMR-CNRS 6112, Nantes Universit\'e, 44322 Nantes cedex 03, France}
\email{mathilde.kervazo@univ-nantes.fr}

\author[orcid=0009-0002-5914-1214]{Owen Lammert}
\affiliation{Waterloo Centre for Astrophysics and Department of Physics and Astronomy, University of Waterloo, Waterloo, Ontario, Canada}
\email{olammert@uwaterloo.ca}

\author[orcid=0000-0002-0747-8862]{Yamila Miguel}
\affiliation{SRON, Space Research Organisation Netherlands, Niels Bohrweg 4, 2333 CA Leiden, The Netherlands}
\affiliation{Leiden Observatory, Leiden University, Leiden, The Netherlands}
\email{ymiguel@strw.leidenuniv.nl}

\author[orcid=0000-0002-5887-1197]{Raymond Pierrehumbert}
\affiliation{University of Oxford, Atmospheric, Oceanic, and Planetary Physics, Oxford, United Kingdom}
\email{raymond.pierrehumbert@physics.ox.ac.uk}

\author[orcid=0000-0001-9504-3174]{Allona Vazan}
\affiliation{Astrophysics Research Center (ARCO), Department of Natural Sciences, The Open University of Israel, Raanana 4353701, Israel}
\email{vazan@openu.ac.il}

\author[orcid=0000-0003-3191-2486]{Joost P. Wardenier}
\affiliation{Weltraumforschung und Planetologie, Physikalisches Institut, University of Bern, Gesellschaftsstrasse 6, 3012 Bern, Switzerland}
\email{joost.wardenier@unibe.ch}

\author[orcid=0000-0003-0562-6750]{Sebastian Zieba}
\affiliation{Center for Astrophysics, Harvard \& Smithsonian, 60 Garden Street, Cambridge, MA 02138, USA}
\email{sebastian.zieba@cfa.harvard.edu}

\collaboration{all}{The Hell of a Survey collaboration}

\begin{abstract}
Ultra-short-period (USP) rocky exoplanets can have dayside temperatures high enough to maintain permanent magma oceans, sitting at the intersection of interior geophysics and atmospheric chemistry. Coupled feedbacks between the molten surface and outgassed atmosphere can sustain or enhance a volatile envelope, while stellar interactions can erode it. Understanding which outcome prevails, and its observable imprint, requires a multi-target approach across planets at different stages of thermal evolution. We present predictions for the five targets of JWST Cycle 4 program 8864: TOI-1807~b, TOI-2260~b, TOI-431~b, TOI-6255~b, and TOI-2431~b. Using the \texttt{PROTEUS} coupled interior--atmosphere framework, we construct a simulation grid and classify outcomes into six categories, defined by the final interior melt state and by whether the planet retains a detectable atmosphere thick enough to redistribute heat to the nightside. For targets retaining a non-negligible volatile envelope, our models predict higher partial pressures for most species when the surface is molten, except for S$_{2}$, whose enhancement in the solid regime suggests it may serve as a tracer of interior melt state. Despite some targets showing outcomes across multiple scenarios, most tend toward a bare-rock end-member, with global melt fraction $\leq$ 20\% and atmospheric retention sensitive to escape efficiency. Our analysis reveals a minimum escape efficiency threshold below which volatile envelopes survive under energy-limited escape, constraining the conditions required for atmosphere survival on irradiated rocky planets. These predictions will guide interpretation of MIRI-LRS phase curve observations and identify which targets and features best discriminate between competing geophysical states.
\end{abstract}


\section{Introduction}
\label{sec:intro}
The early evolution of rocky planets is marked by a fully molten phase, characterized by the presence of a global magma ocean \citep{elkins-tanton_magma_2003}. The energy deposited during accretion and collisions is sufficient to melt the silicate mantle \citep{elkins-tanton_magma_2012, johansen_forming_2017, raymond_impact-driven_2020, batygin_formation_2023}. The subsequent cooling and crystallization of this molten stage sets the initial conditions for the structural and thermal evolution of the planet -- namely, the segregation of an iron core, the outgassing and build-up of the atmosphere, and the long-term volatile recycling that governs climate evolution over geological timescales \citep{elkins-tanton_linked_2008, lebrun_thermal_2013, hamano_emergence_2013, bower_linking_2019, bower_retention_2022, salvador_magma_2023, Lichtenberg2023ASPC, nicholls_magma_2024, boukare_2025}. Over the past decade, exoplanet surveys have revealed a growing population of highly irradiated rocky planets whose extreme stellar irradiation sustains long-lived magma oceans, preventing complete solidification.

Therefore, these rocky worlds, historically referred to as `Hot Earths'' or `Super Hot Earths'' in the earlier literature \citep{leger_2011, winn_2018}, are now more commonly described as ``lava worlds.'' They are ultra-short-period (USP) rocky planets, with orbital periods of only a few hours \citep{leger_2009, sahu_2006, adams_2016, batalha_2011, goyal_2025}, and orbit so close to their host stars that they receive irradiation levels orders of magnitude higher than those experienced by Earth. Their high bulk densities are consistent with predominantly rocky compositions \citep{dai_2019}.
 Under such extreme conditions, synchronous rotation due to intense tidally-forced spin-down \citep{leger_2009,Farhat2025ApJ979133F} allows the dayside temperatures to be hot enough to melt the silicate mantle \citep{dai_2019} and any primordial hydrogen envelope is expected to be rapidly lost during the early stages of evolution \citep{wordsworth_atmospheres_2022,sanchis_2014,lundkvist_2016,lopez_2017}, leaving behind several possible end-states: bare rock with a highly reflective surface, a rock-vapour envelope sustained by vapour-pressure equilibrium with a molten dayside \citep{miguel_2011, schaefer_2009}, or a secondary volatile atmosphere outgassed from the interior over geological timescales \citep{Lichtenberg2025TrGeo,Lichtenberg2025arXiv}. Each of these outcomes is the product of a distinct interior thermal history and initial volatile budget, shaped in turn by the stellar type that sets the planet's irradiative and thermal environment, and each leaves a characteristically different imprint on the planet's observable properties. \citep{hu_2012, demory_2016, kreidberg_absence_2019, dai_2019, whittaker_2022, mercier_2022, crossfield_2022, zhang_2024, dai_2024}. The fundamental question is therefore which of these states a given USP lava exoplanet currently occupies and whether its thermal emission, albedo, and day-to-night temperature contrast carry sufficient information to uniquely identify it and, by extension, constrain the geophysical processes that shaped it.

Addressing this requires a theoretical foundation rooted in the coupled evolution of a planet's interior and atmosphere, from the fully molten stage through volatile outgassing and eventual solidification \citep{elkins-tanton_linked_2008}. As volatiles are released from the melt, the resulting atmosphere can generate a considerable greenhouse effect, trapping outgoing radiation and maintaining surface temperatures above the solidus, the pressure- and composition-dependent threshold for melting. This feedback loop between outgassing and warming can substantially extend the lifetimes of magma oceans \citep{abe_1985,abe_1988,schaefer_predictions_2016,salvador_relative_2017,lichtenberg_vertically_2021}. Both molten and solid surfaces produce different atmospheric compositions \citep{hamano_2015, schaefer_2016, shorttle_2024, nicholls_magma_2024, boukare_2025, lichtenberg_2025} because outgassed volatile compounds have different affinities for dissolution into silicate melts and different redox behaviours \citep{Suer2023FrEaS}, offering a unique opportunity to probe the interior state of rocky exoplanets.

Recent observations have begun to reveal the complexity of this picture. For instance, TOI-431 b \citep{osborn_2021}, a USP lava planet with an equilibrium temperature of 1862 $\pm$42 K \citep{osborn_2021} sufficient to maintain a dayside magma ocean, was observed by Spitzer in thermal emission \citep{monaghan_2025}. The resulting constraints are consistent with either a bare silicate surface with a high Bond albedo or a tenuous atmosphere, leaving the two scenarios observationally degenerate. The same ambiguity is broken in the case of HD 3167 b, for which a JWST Mid-Infrared Instrument Low-Resolution Spectroscopy (MIRI-LRS) secondary-eclipse observation yielded an eclipse depth more than 5$\sigma$ shallower than expected for a dark, hot bare-rock surface \citep{coy_2026}. The reduced dayside emission is best explained by an atmosphere that cools the planet through reflection and/or heat redistribution, providing evidence that at least some highly irradiated rocky planets can sustain atmospheres capable of altering their observable thermal emission.

Evidence for atmospheric retention on such worlds is further strengthened by observations of 55 Cnc e. A persistent volatile envelope was detected around the planet, with a composition inconsistent with either bare rock or silicate vapour and likely dominated by CO or CO$_2$ \citep{hu_secondary_2024}. Although occultation depths show substantial variability between epochs \citep{patel_2024}, the observations provided the first direct evidence that secondary atmospheres can survive on highly irradiated rocky planets. Re-analysis of the same JWST dataset using cross-correlation spectroscopy further demonstrated that these atmospheres can be compositionally constrained: \citet{snellen_2026} reported a robust detection of CO emission in one of five observing epochs, with tentative detections in two additional epochs, while placing stringent limits on the CO$_2$/CO abundance ratio and finding CO$_2$ to be at least three orders of magnitude less abundant than CO.

Atmospheric effects are also apparent in the thermal emission of other lava worlds, even when the atmosphere is not directly detected spectroscopically. The dayside emission of TOI-561b \citep{lacedelli_2022}, for example, requires efficient heat redistribution across the planet, favouring the presence of a substantial atmosphere over an exposed rocky surface \citep{teske_thick_2025}. Importantly, the host star TOI-561 is metal-poor and $\alpha$-enhanced ([Fe/H]~$=-0.40\pm0.12$, [$\alpha$/Fe]~$=0.23\pm0.04$; \citealt{lacedelli_2022}), suggesting that the planet-forming environment was depleted in iron relative to $\alpha$-elements. Such a composition may have influenced the planet's volatile-to-refractory element budget and interior composition.

Theoretical works predict that the interior thermal evolution of USP lava planets converges toward one of two stable end-member states, either a hot or cold internal scenario \citep{boukare_2025,meier_2023}. In the hot end-member state, the mantle remains fully molten, and vigorous turbulent convection within the magma ocean homogenizes the interior, driving efficient heat transport from the irradiated dayside to the nightside through the mantle itself; this efficient heat transfer sustains nightside surface temperatures of $\sim$1500 K. The overlying atmosphere reflects the bulk silicate composition of the planet. In the cold end-member, the interior has largely solidified, leaving a shallow dayside magma ocean confined by a viscous solid mantle; solid-state convection is insufficient to transport significant heat to the nightside, and nightside temperatures fall to the order of 100 K. The atmosphere in this scenario is depleted of volatile species \citep{boukare_2025}.  

These global scenarios may also be complicated by hemispheric asymmetry: 2D geodynamical simulations show tidally locked lava planets can develop distinct dayside/nightside tectonic and convective regimes \citep{meier_hemispheric_2021,meier_2023,Meier2024JGRE12908491M, herath_2024}. On planets with a molten dayside and a solid or mushy (partially molten) nightside, atmospheric replenishment may proceed through fundamentally different mechanisms on each hemisphere. Core-mantle boundary superplumes preferentially feed the dayside magma ocean with volatiles and reaction products, enriching the overlying atmosphere in CO, CO$_2$, H$_2$, and H$_2$O, whereas the nightside relies solely on volcanic outgassing from the solid mantle. The resulting hemispheric asymmetry could produce compositionally distinct dayside and nightside atmospheres, as proposed for 55 Cnc e \citep{meier_2023}. Nightside temperature is therefore a key observable for distinguishing these interior states. Moreover, in the likely event that lava planets do not form in situ but rather migrated to their current short orbital periods, then their thermal state need not evolve monotonically, e.g., \citet{herath_2026} predict that migrating lava planets undergo three distinct epochs in which they are fully molten.  Together, these predictions show that irradiated rocky planets span diverse interior-atmosphere outcomes, requiring a self-consistent framework coupling interior thermal evolution to atmospheric composition and observable emission.

These questions will be directly addressed by JWST Cycle 4 program 8864, the ``Hell of a Survey'' \citep{dang_proposal_2025}, a dedicated survey designed to characterize the thermal and atmospheric states of five USP lava planet targets: TOI-1807 b \citep{nardiello_2022, macdougall_2023, barber_2025}, TOI-2260 b \citep{giacalone_2022}, TOI-431 b \citep{osborn_2021}, TOI-6255 b \citep{dai_2024}, and TOI-2431 b \citep{tas_2026}. Each target will be observed with JWST/MIRI-LRS in slitless prism mode (SLITLESSPRISM subarray, FAST readout), following target acquisition with the F1500W filter. The resulting full-orbit thermal phase curves provide continuous coverage of both the dayside and nightside emission, enabling direct measurements of the day-night temperature contrast, heat redistribution efficiency, and the presence or absence of an atmosphere. The five targets were selected to span a broad range of system ages (Section~\ref{sec:targets}). This age-sequence design is a defining feature of the program: rather than treating each planet as an independent case study, the survey is constructed to trace the evolutionary trajectory of USP planets as a population, sampling distinct stages of interior cooling and atmospheric evolution within a homogeneous dataset. In this context, stellar age provides a critical additional constraint: although the planets remain persistently irradiated by their host stars, the high-energy (XUV) output of the host declines by orders of magnitude over Gyr timescales, progressively allowing the interior to cool down towards the solidification. As a result, for the oldest targets in the sample, this decline should have driven the interior toward the cold end-member, whereas the youngest targets, still experiencing comparatively high XUV irradiation, are expected to remain closer to the hot end-member, having had comparatively little time to cool. \citep{boukare_2025,Farhat2025ApJ979133F}.

Beyond the interior thermal state, program 8864 \citep{dang_proposal_2025} also aims to characterize the outgassed atmospheres of these targets, if present, thereby directly benchmarking observable atmospheric properties against the geophysical state of the underlying interior and providing empirical constraints for coupled interior–atmosphere evolution models.

In this context, predictive physical modelling is not merely complementary to observation but essential to it. The spectral degeneracies described above where bare surfaces, rock-vapour atmospheres, and secondary volatile envelopes can produce similar broadband emission signatures mean that data interpretation cannot proceed without a theoretical framework that maps geophysical states to observable quantities and thereby allows observations to be traced back to the states that produced them. By generating physically motivated predictions for nightside temperatures, emission spectra, and atmospheric compositions across a range of planetary ages and irradiation levels, such models will allow quantitative constraints on interior melt fraction, initial volatile inventory, and evolutionary history.

The paper is structured as follows: Section~\ref{sec:methods} presents the model setup, target properties, and simulation grid; Section~\ref{sec:results} presents the sensitivity analysis and detectability predictions for each planet; Section~\ref{sec:discussion} discusses the implications for the ``Hell of a Survey'' observations and for atmospheric retention on rocky exoplanets more broadly; and Section~\ref{sec:conclusions} provides our conclusions.

\section{Methods}
\label{sec:methods}

In this work, we model the thermal and compositional evolution of each target using the \texttt{PROTEUS} framework\footnote{\url{https://proteus-framework.org}} (v25.10.15; \citealt{lichtenberg_vertically_2021, nicholls_magma_2024, nicholls_self-limited_2025, nicholls_volatile-rich_2025}), which self-consistently couples magma-ocean cooling, volatile outgassing, and atmospheric chemistry over time. For each planet, observed system properties including mass, radius, semi-major axis, and stellar mass fix the initial conditions of the models, while uncertainties in the initial thermal state, volatile inventory, interior structure, and atmospheric escape are explored through a discrete, multidimensional grid of model realizations. The grid spans a broad range of physically motivated values for each free parameter; the specific parameter values and sampling intervals are described in Section~\ref{sec:params} and summarised in Table~\ref{tab:params}. Throughout, we use the fraction of grid realizations that converge to quantify how extensively that outcome is represented within the explored parameter space, rather than interpreting it as a formal probability. The resulting evolutionary tracks characterize the melt state, surface pressure, and atmospheric composition (if present) as a function of time, allowing us to quantify which geophysical states are consistent with detectable atmospheric signatures and which can be ruled out by their absence. We define a detectability metric that maps these outcomes onto predicted MIRI-LRS observables, enabling direct comparison between model predictions and the observations obtained through JWST Program 8864 \citep{dang_proposal_2025}.

\subsection{Target Planets}
\label{sec:targets}

\begin{figure}[h]
    \centering
    \includegraphics[width=\linewidth]{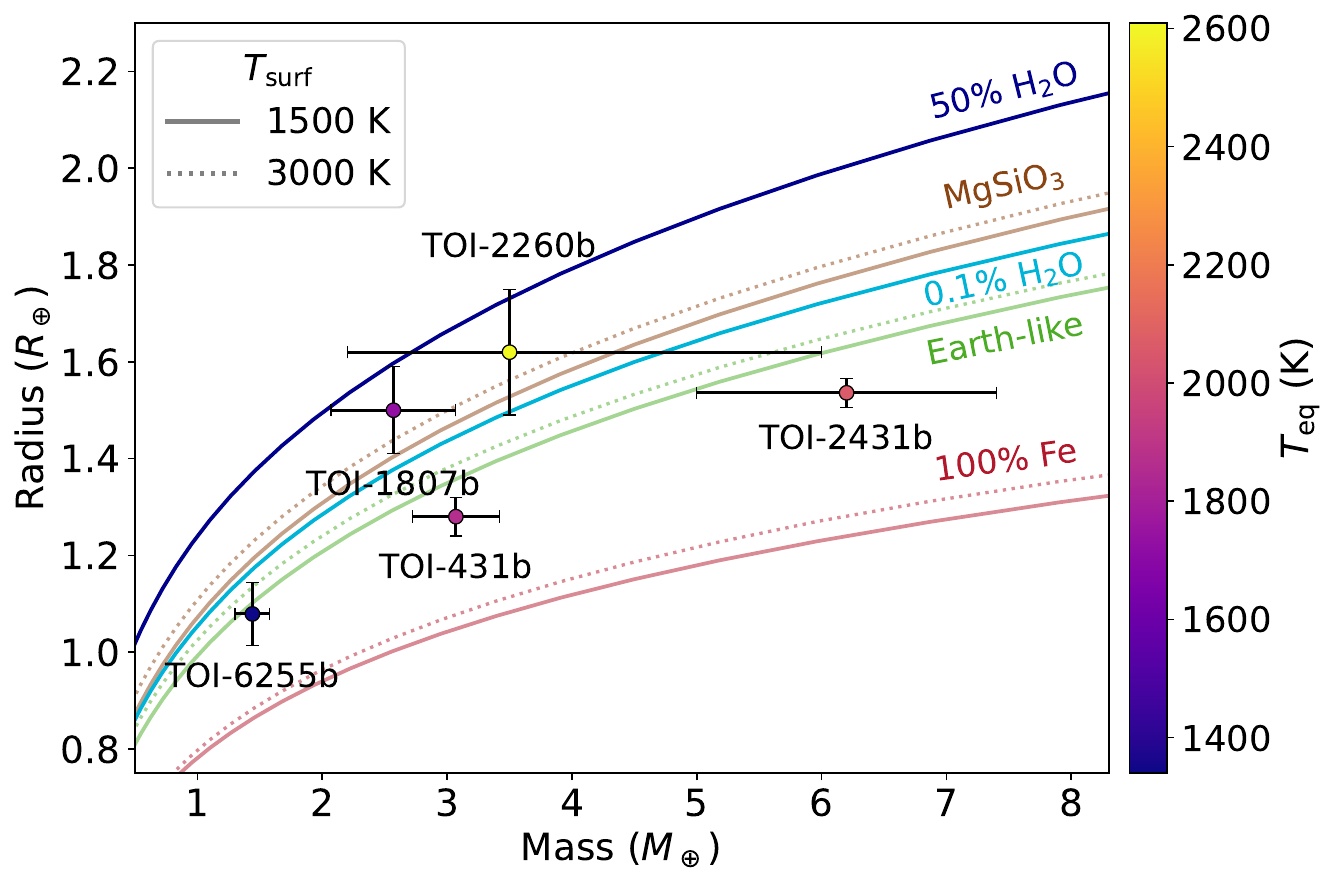}
    \caption{Mass–radius diagram of the rocky exoplanet targets in this study. Colored circles indicate each planet, with the color bar representing equilibrium temperature $T_\mathrm{eq} (K)$. Theoretical temperature-dependent-mass–radius curves from \citet{attia_2026} are shown as dashed and continuous lines for reference surface temperatures and compositions. Error bars reflect reported uncertainties in planetary mass and radius for each target. Note that the mass adopted for TOI-2260~b is not an RV measurement but a model-predicted value \citep{giacalone_2022}.}
    \label{fig:M-R}
\end{figure}

The targets of JWST GO Program 8864 span a wide range of ages, enabling exploration of planets at different stages of thermal evolution. TOI-1807 b and TOI-2260 b are relatively young systems (<1 Gyr), when atmospheric escape and outgassing may still be shaping their present-day atmospheres. In contrast, TOI-431 b, TOI-6255 b, and TOI-2431 b are older systems (>2 Gyr), with interiors expected to have undergone substantial cooling and solidification, placing them closer to their long-term thermal state \citep{dang_proposal_2025}. All five targets are ultra-short-period planets, with orbital periods of less than 0.6 days, hosted by stars spanning late K to early M spectral types (TOI-1807: K3V; TOI-2260: late G; TOI-431: early K; TOI-6255: M dwarf; TOI-2431: K7V), and with Earth-like bulk densities, to avoid the ambiguity introduced by thick primordial envelopes and to focus on signals originating from outgassed or rock-vapour atmospheres. Dayside equilibrium temperatures are above $\sim$1000~K across the sample, adopted from the literature and computed assuming zero Bond albedo and full day-night heat redistribution (except for TOI-6255 b, for which an albedo of 0.1 was assumed; \citealt{dai_2024}), are sufficient to sustain molten, evaporating surfaces, providing a direct window into planetary interiors.  
The physical and orbital properties of each target drawn from the literature are listed in Table~\ref{tab:targets}, and their location in mass-radius diagram is shown in Fig.~\ref{fig:M-R}.

We note that the mass listed for TOI-2260 b in Table~\ref{tab:targets} is not derived from radial-velocity observations, as no such measurements have been published for this system. Instead, \citet{giacalone_2022} forecast this value using a probabilistic mass-radius relation based on \citet{chen_2017}, and explicitly caution against treating it as an actual mass. We adopt this predicted value only as a working estimate and return to the implications of this uncertainty in Section~\ref{sec:discussion}. Of the five targets, TOI-431 b stands apart as it has already been observed in thermal emission \citep{monaghan_2025} with Spitzer, offering an opportunity to test our theoretical predictions. All observations for the program are expected to be completed by the end of August 2026. In addition, TOI-431 b and TOI-1807 b will have MIRI- LRS secondary eclipses observed under program 4818 \citep{mansfield_jwst}, complementing the phase curves obtained in this program.

\begin{table*}[t!]
\centering
\caption{Observed properties of the five JWST Program 8864 target planets.}
\label{tab:targets}
\begin{tabular}{lcccccccl}
\hline\hline
Target & $M_\oplus$ & $R_\oplus$ & $T_\mathrm{eq}$ [K] & Period [days] & $M_\star$ [$M_\odot$] & $T_\star$ [K] & Age [Gyr] & Reference \\
\hline
1807 b  & $2.57\pm0.50$       & $1.5\pm0.09$    & $1730\pm28$ & 0.549374 & $0.8\pm0.03$    & $4730\pm75$  & $0.210 \pm0.02$ & \citenum{nardiello_2022, macdougall_2023, barber_2025}\\
2260 b& $3.5^{+2.5}_{-1.3}$$^{*}$& $1.62\pm0.13$   & $2609\pm86$ & 0.352    & $0.99\pm0.04$   & $5534\pm100$ & $0.32\pm0.10$       & \citenum{giacalone_2022}\\
431 b   & $3.07\pm0.35$       & $1.28\pm0.04$   & $1862\pm42$ & 0.49     & $0.78\pm0.07$   & $4850\pm75$  & $5.1\pm0.6$         & \citenum{osborn_2021}\\
6255 b  & $1.44\pm0.14$       & $1.079\pm0.065$ & $1340\pm60$ & 0.238    & $0.353\pm0.015$ & $3421\pm70$  & $6.0\pm2.0$         & \citenum{dai_2024}\\
2431 b  & $6.2\pm1.2$         & $1.536\pm0.03$  & $2063\pm30$ & 0.224    & $0.661\pm0.02$  & $4109\pm28$  & $2.0^{+9.0}_{-1.7}$ & \citenum{tas_2026}\\
\hline
\multicolumn{9}{l}{\footnotesize $^{*}$Model-predicted, not an RV measurement; see Section~\ref{sec:targets} for details.}\\
\end{tabular}
\end{table*}

\subsection{The PROTEUS Framework}
\label{sec:proteus}

We model the coupled interior-atmosphere evolution of each planet in the sample using \texttt{PROTEUS}, a modular 1D framework that tracks the thermal and compositional history of a planet from an initially molten state through magma ocean crystallization and volatile outgassing \citep{lichtenberg_vertically_2021, nicholls_magma_2024, nicholls_self-limited_2025, nicholls_volatile-rich_2025}. The code couples six stand-alone modules that simulate the planet's interior \citep{bower_numerical_2018, bower_retention_2022}, tidal heating \citep{nicholls_self-limited_2025, van_Dijk_2026}, volatile outgassing \citep{bower_retention_2022, nicholls_convective_2025}, radiative-convective atmospheric evolution \citep{nicholls_agni_2025}, atmospheric escape \citep{postolec_2026}, and stellar evolution \citep{johnstone_2021}. At each timestep, the modules exchange energy, mass, and compositional fluxes through shared boundary conditions until a self-consistent converged solution is obtained. We refer the reader to \citet{lichtenberg_vertically_2021, nicholls_magma_2024, nicholls_volatile-rich_2025, nicholls_thesis_2026} for a comprehensive description of the framework.

The planetary interior is evolved using the module \texttt{SPIDER} (Simulating Planetary Interior Dynamics with Extreme Rheology) \citep{bower_numerical_2018, bower_linking_2019}, which solves the entropy-based energy conservation equation on a 1D radial grid, accounting for convection via mixing length theory, conduction, convective mixing, and gravitational settling. Phase transitions are described through MgSiO$_3$ melt and solid equations of state \citep{wolf_equation_2018, mosenfelder_mgsio3_2009}, with melting curves from \citet{andrault_solidus_2011} and \citet{hamano_emergence_2013}. Tidal heating is computed using the viscoelastic model \texttt{LovePy} \citep{hay_2019}, which accounts for the radially varying tidal response of the planet \citep{sabadini_2016}. The mantle is modeled as a Maxwell viscoelastic solid with melt-dependent rheological properties \citep{kervazo_2021}. As the melt fraction increases, the associated decrease in viscosity and shear modulus modifies the tidal response, with dissipation enhanced at intermediate melt fractions and strongly suppressed above the critical melt fraction, $f_{\rm crit}\approx0.3$ \citep{arzi_1978}. The geophysical evolution of the metallic core is not modeled explicitly. Instead, the core-mantle boundary (CMB) is assumed to be thermally coupled to the mantle through a heat flux determined by energy balance \citep{bower_numerical_2018}. The bulk C-H-O-N-S inventory, initially assumed to reside entirely in the atmosphere, is partitioned between the melt and atmosphere following the outgassing module \texttt{CALLIOPE} \footnote{\url{https://proteus-framework.org/CALLIOPE/}} \citep{bower_retention_2022, nicholls_convective_2025, nicholls_thesis_2026}. Volatile solubilities depend on the surface temperature, melt fraction, and prescribed oxygen fugacity \fotwo, expressed relative to the iron–wüstite (IW) buffer as the $\Delta$IW offset \citep{frost_2008, hirschmann_magma_2012, cottrell_2025}. Sulphur follows the sulphide-saturated mafic-melt law of \citet{gaillard_redox_2022}, in which solubility increases with magma temperature and towards more reducing conditions (lower \fotwo), where sulphur dissolves as sulphide (S$^{2-}$). Atmospheric escape is modelled under an energy-limited formulation, in which the mass-loss rate scales with the incident XUV flux from the evolving host star, according to Equation~\ref{eq:escape} \citep{watson_1981, erkaev_2007, lopez_2013}:
\begin{equation}
\label{eq:escape}
    \dot{M}_{\mathrm{EL}} = \frac{\eta \pi R_{\mathrm{XUV}}^3 F_{\mathrm{XUV}}}{G M_p K_{\mathrm{tide}}},
\end{equation}
where $\dot{M}_{\mathrm{EL}}$ is the mass-loss rate in kg\,s$^{-1}$, $R_{\mathrm{XUV}}$ is the radius at which XUV radiation affects the atmosphere, $F_{\mathrm{XUV}}$ is the stellar XUV flux received by the planet in W\,m$^{-2}$, $G$ is the gravitational constant in N\,m$^{2}$\,kg$^{-2}$, $M_p$ is the planetary mass in kg, and $K_{\mathrm{tide}}$ is the tidal correction factor, which we set to zero here. The escape efficiency $\eta$ is a dimensionless parameter in the range $(0,1]$ representing the fraction of incident XUV energy converted into work against gravity to drive the outflow \citep{postolec_2026}. In this case, treated as a free parameter in our grid (see Sect.~\ref{sec:params}). Each simulation is integrated forward to the estimated stellar age of each system, which we adopt as a proxy for the current planetary state. We do not formally propagate the uncertainties on stellar age through our simulations; rather, this approach is justified by the fact that the key phases of magma ocean evolution, including rapid crystallization and convergence towards long-lived equilibrium states, occur within the first $10$-$100$~Myr \citep{nicholls_magma_2024}, well within even the lower bound of the age uncertainty for most of the targets (Table~\ref{tab:targets}), such that the planet formation stage can be regarded as instantaneous relative to the total evolution time.

\subsection{Parameter Space}
\label{sec:params}

The initial volatile budgets and interior compositions of rocky exoplanets are poorly constrained, and are expected to vary significantly across formation environments \citep{Krijt2023ASPC,Drazkowska2023ASPC}. To account for this uncertainty, we construct a grid spanning a wide range of initial conditions for each of the five targets. The total hydrogen inventory is 
expressed in Earth ocean units and the bulk C/H ratio is sampled across values that bracket refractory and volatile-rich compositions \citep{wang_elemental_2018}. In contrast, the sulfur and nitrogen inventories are not varied independently: they are fixed relative to hydrogen at mass ratios derived from Earth's primitive mantle \citep{wang_elemental_2018}. The mantle redox state \fotwo expressed in relative terms of $\Delta$IW is explored over a seven-point range covering highly reduced interiors, analogous to Mercury \citep{cartier_role_2019}, through to oxidised mantles comparable to present-day Earth \citep{stagno_2020}. $\Delta$IW is treated as an independent model input, setting the oxygen budget available for exchange between mantle and atmosphere. This is motivated by the recognition that mineralogical heterogeneity alone can drive $f\mathrm{O}_2$ variations exceeding several orders of magnitude \citep{guimond_mineralogical_2023}.

Given that several planets in our sample have bulk densities inconsistent with a canonical Earth-like iron-to-rock ratio at their observed masses and radii (Fig.~\ref{fig:M-R}), we treat the core radius fraction (CRF) as an additional free parameter. Varying this quantity allows us to probe a wider range of interior structures and its downstream effect on mantle mass, outgassing capacity, and interior pressure gradients. As mentioned above, the atmospheric escape efficiency $\eta$ is included as a free parameter, both reflecting the substantial uncertainties in XUV-driven escape for highly irradiated super-Earths and identifying the minimum escape efficiency threshold below which a planet is capable of retaining a substantial atmosphere. Given the diversity in irradiation, mass, and age in our target sample, we investigate whether this threshold is consistent across those five planets or otherwise varies depending on their individual system properties. The set of values used for each of these parameters is summarised in Table~\ref{tab:params}.

\begin{table}[h]
\centering
\begin{tabular}{ll}
\hline\hline
Parameter & Values \\
\hline
H inventory [Earth oceans]        & $3,\,10,\,100$ \\
$\Delta$IW (\fotwo)               & $-3,\,-2,\,-1,\,0,\,+1,\,+2,\,+3$ \\
Bulk C/H mass ratio               & $0.1,\,2.0,\,8.0$ \\
Bulk S/H mass ratio               & $2.0$ \\
core radius fraction                & $0.4,\,0.5,\,0.6,\,0.7,\,0.8$ \\
Escape efficiency                 & $10^{-4},\,10^{-3},\,10^{-2}$ \\
\hline
\end{tabular}
\caption{Overview of the parameter space explored in this study.}
\label{tab:params}
\end{table}

\subsection{Regime classification}
\label{sec:filtering}
Out of the 945 simulations per planet (the full grid obtained from combining all explored parameter dimensions) we focus our analysis on those whose final state is consistent with the observed system properties. Specifically, we retain simulations for which the predicted bulk density at the planet's age (taken to be the stellar age; Table~\ref{tab:targets}) is consistent, within 1$\sigma$, with the density implied by the observed planetary mass and radius (Table~\ref{tab:targets}); the age itself is held fixed and does not contribute to this uncertainty range. We refer to these as density-consistent cases, and all subsequent analysis is performed on this subset. Each density-consistent simulation is then classified according to its final interior and atmospheric state. The interior melt state is defined based on the global melt fraction $\phi$ at the end of the simulation:
\begin{equation}
\label{interior_regimes}
\text{Interior state} =
\begin{cases}
\text{Molten,}        & \phi \geq 0.8 \\
\text{Partial melt,}  & 0.2 < \phi < 0.8 \\
\text{Solid,}         & \phi \leq 0.2
\end{cases}
\tag{I}
\end{equation}

Classification~\ref{interior_regimes} defines the bulk interior state based on the global melt fraction $\phi$; a planet labelled "solid" ($\phi \leq 0.2$) under this definition does not preclude localized dayside lava pools sustained by the intense stellar flux, even as the bulk interior has crystallized. Further down below we classify this regime as "bare rock" cases for simplicity, but would like to note that such shallow, spatially confined melt features fall below the spatial resolution of our models, which track global-scale interior evolution rather than resolving these localized dayside structures, and their associated atmospheric consequences (e.g., silicate vapor). The specific melt-fraction thresholds adopted for each regime are chosen to reflect the qualitative behaviour of the relative viscosity as a function of solid volume fraction based on \citep{costa_2009}, which starts to transition between solid-like and liquid-like rheology around these values.
Classification \ref{eq:atmosphere_regime} defines the atmospheric state according to the final surface pressure, $P_\mathrm{surf}$. We adopt a threshold of $0.1$~bar, below which the atmosphere is too tenuous to produce a detectable day-night heat redistribution signature in photometric phase curves \citep{ito_2021, kreidberg_absence_2019}.This is particularly relevant for the highly irradiated planets considered here, where even a $\sim$0.1~bar atmosphere may be difficult to sustain against atmospheric escape. We note this threshold reflects redistribution detectability specifically, rather than spectroscopic detectability; a $\sim$0.1~bar atmosphere could still yield observable spectral features in thermal emission depending on composition and the vertical temperature gradient \citep{hu_secondary_2024, piaulet_2025}

\begin{equation}
\label{eq:atmosphere_regime}
\text{Atmospheric state} =
\begin{cases}
\text{Atmosphere,} & P_\mathrm{surf} > 0.1~\text{bar} \\
\text{Bare/thin,}  & P_\mathrm{surf} \leq 0.1~\text{bar}
\end{cases}
\tag{II}
\end{equation}
Combining both classifications yields six possible outcome categories: molten, partially molten, or solid interior, each paired with the presence or absence of a thick atmosphere. Finally, to quantify which atmospheric species are most likely detectable, we define a detectability metric based on the density-consistent simulations that host a substantial atmosphere ($P_\mathrm{surf} > 0.1$~bar). For each planet, we compute the fraction of simulations falling into each of the six outcome categories defined above. Within the atmosphere-bearing cases, we then catalogue the partial pressure of each species and focus our analysis on those exceeding $0.1$~bar, below this threshold which we consider their contribution negligible for detection of an efficient heat redistribution. This allows us to identify, for each interior regime separately (molten, mushy, and solid), which species are present at detectable levels and how their partial pressures are distributed across the parameter space. The result is a per-planet, per-regime summary of the most dominant atmospheric constituents, which we use to assess the detectability of each target with MIRI-LRS and to identify the spectral signatures associated between interior states.

\section{Results}
\label{sec:results}

In this section, we present the outcomes of our simulation grid for each of the five targets, structured around the three questions motivating this survey: whether the planets are currently molten (Sect.~\ref{sec:results_melt}), what atmospheric states they are likely to host (Sect.~\ref{sec:results_atm}), and how these predictions translate into detectable observational signatures (Sect.~\ref{sec:results_detect}). For each question, we examine trends across the density-consistent subset of simulations, isolating the influence of individual free parameters before turning to planet-specific results.

\subsection{Are the Planets Currently Molten?}
\label{sec:results_melt}
We investigate the sensitivity of the predicted interior melt state to each of the free parameters in our grid across the density-consistent simulations for all five targets. All planets in our sample are subject to extreme stellar irradiation, with equilibrium temperatures (Table~\ref{tab:targets}) well above the silicate solidus at the surface ($\sim$1373~K at 0~GPa; \citealt{katz_2003}), making them prime candidates for hosting a present-day magma ocean on the day-side. For each explored parameter, we collapse the remaining dimensions of the grid and examine the resulting distribution of melt fractions and atmospheric surface pressures, allowing trends driven by individual variables to emerge independently of the others. The mean atmospheric composition is computed in the same way, marginalising over all other free parameters to isolate the contribution of each driver.

Before presenting the results for the density-consistent simulations, we briefly consider the simulations excluded by the density filter described in Section.~\ref{sec:filtering}. Out of the 945 simulations in each planet's grid, more than 500 fail this filter for TOI-431 b and TOI-2431 b alone. Some of this is due to numerical instabilities caused by the extreme atmospheric temperatures of these planets, but structural factors are just as important: both planets are observed to have  high bulk densities, and the density-inconsistent cases are concentrated at low-to-intermediate core radius fractions (CRF~$\leq 0.6$ Fig \ref{fig:regimes} panel C5 and Fig \ref{toi2431} panel C5). This pattern shows that an Earth-like Fe/Si ratio cannot reproduce the observed mass and radius of these two planets, regardless of the assumed thermal or volatile state, and instead suggests that their bulk composition is iron-enriched from formation, consistent with their position below the Earth-like composition curve in the mass–radius diagram (Figure~\ref{fig:M-R}).

For the remaining planets, a different but related trend emerges: for small cores, the volatile inventory dictates the mantle melt state, with low volatile inventories tending to produce solid mantles and high volatile inventories instead producing the largest fully molten mantles. This is because a large volatile inventory enhances the atmospheric blanketing effect, trapping interior heat and keeping the mantle hot enough to remain fully molten; once thermal expansion is taken into account, this hot, molten mantle is inflated relative to a cooler, solid one, lowering the bulk density below the observed range, which is precisely why these large mantles, volatile-rich solutions are excluded by the density filter. This indicates that these planets did not form with a configuration that paired a substantial rocky mantle with a correspondingly large volatile inventory, in agreement with \citet{krissansen-totton_predictions_2022}. We discuss the outcome for each planet individually below.

\subsubsection{TOI-1807 b}

\begin{figure*}[t!]
    \centering
    \includegraphics[width=\linewidth, height=0.75\textheight, keepaspectratio]{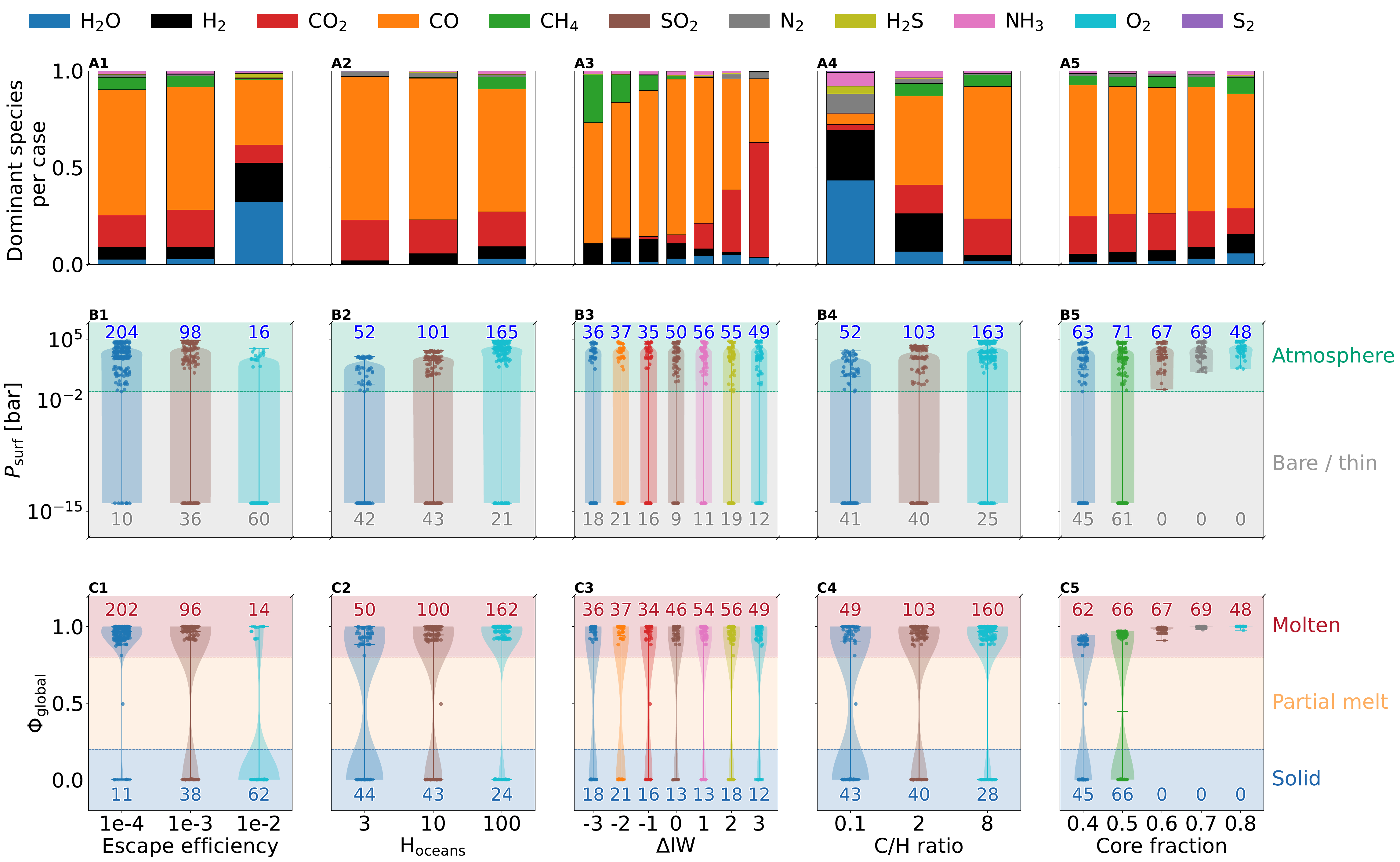}
    \caption{Summary of theoretical outcomes for TOI-1807 b across the 398 density-consistent cases (out of 945 simulations), as a function of escape efficiency, initial water content \hoceans, oxidation state \fotwo, C/H ratio, and core radius fraction (columns). First row: median atmospheric composition per bin, computed from per-case proportional volume mixing ratios ($x_i = p_i/P_\mathrm{surf}$), so each simulation contributes equally regardless of its total pressure. Second row: distribution of absolute final surface pressure $P_\mathrm{surf}$; numbers give the count of cases with a substantial (top) and bare/thin (bottom) atmosphere, and the dashed line marks $P_\mathrm{surf} = 0.1$\,bar. Third row: distribution of global melt fraction $\Phi_\mathrm{global}$; numbers give the count of molten and solid cases, and shaded regions denote the molten ($\Phi_\mathrm{global} > 0.8$, red), partial melt, and solid ($\Phi_\mathrm{global} < 0.2$, blue) regimes. Violin widths reflect the density of cases, with individual cases overplotted as points.}
    \label{fig:toi1807}
\end{figure*}
Among the 424 density-consistent simulations for TOI-1807 b, the results reveal two dominant outcomes: a molten planet hosting a substantial atmosphere, and a bare rock with a solid interior. For instance, Panel C1 of Figure \ref{fig:toi1807} shows how the escape efficiency $\eta$ has a strong influence on the interior state. At $\eta = 10^{-2}$, the planet is predominantly solid, while decreasing $\eta$ progressively shifts the distribution towards higher melt fractions and atmosphere-bearing solutions. The median atmospheric composition across the three values of $\eta$ shown in Panel A1, exhibits atmospheres consistently dominated by CO$_2$, CO, CH$_4$, and H$_2$, with \water being more pronounced in $\eta = 10^{-2}$, consistent with the outgassing in late stages of crystallisation.  Panel C2, shows the effect of a larger hydrogen inventory favouring a molten interior, consistent with previous results \citep{nicholls_magma_2024, lichtenberg_vertically_2021}. This is, a higher \water\ outgassing drives a greenhouse feedback that sustains elevated mantle temperatures, maintaining the planet in a molten state. The median atmospheric composition is broadly similar across the three H inventories, with an enhanced H$_2$ contribution at 100~\hoceans\ relative to the more water-poor cases. Panel C3 shows how more oxidised conditions (higher \fotwo) correlate with a greater prevalence of molten solutions and a higher fraction of atmosphere-bearing cases. The median atmospheric composition shows a systematic increase in \cotwo\ towards more oxidised values of \fotwo, consistent with thermochemical predictions for oxidised outgassing \citep{bower_retention_2022}. For panel C4, higher C/H ratios are associated with a greater frequency of retaining an atmosphere. At low C/H, the atmosphere is dominated by \water\ and H$_2$, while high C/H cases show a stronger contribution from CO. The overall melt state distribution shifts moderately towards atmosphere-bearing solutions as C/H increases, presumably reflecting differences in melt solubility between H$_2$/H$_2$O and carbon-bearing species \citep{sossi_2020}. Finally, panel C5 shows how TOI-1807 b is consistent with a range of interior structures: solutions with a small core (CRF~$=0.4$) include both solid and molten outcomes, whereas a large core (CRF~$=0.8$) yields exclusively molten solutions within the density-consistent subset. A plausible explanation is that the thinner mantle associated with a large CRF results in a strong temperature gradient that does not fall below the liquidus, especially when heat loss is inhibited by the insulating presence of an atmosphere, which is the case in Panel B5. Despite this structural diversity, the median atmospheric compositions across core radius fractions are largely indistinguishable, meaning that core size affects whether an atmosphere is retained and the planet's thermal state, rather than the chemical character of the outgassed atmosphere.

\subsubsection{TOI-431 b}

\begin{figure*}[t!]
    \centering
    \includegraphics[width=\linewidth]{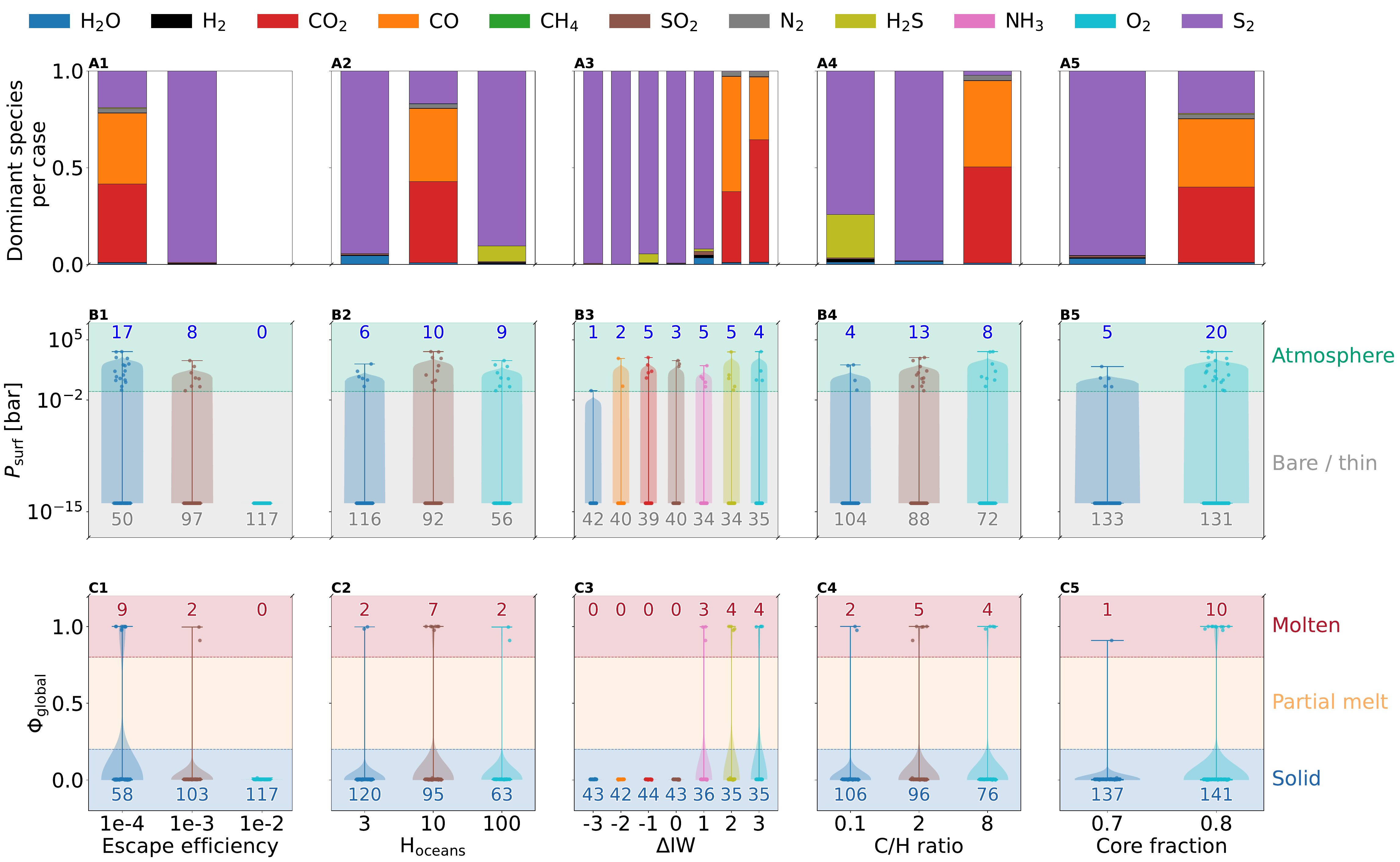}
    \caption{Summary of theoretical outcomes for TOI-431 b across the 289 density-consistent cases, as a function of five input parameters: escape efficiency, initial water content \hoceans, oxidation state \fotwo, C/H ratio, and core fraction.}
    \label{fig:regimes}
\end{figure*}

TOI-431 b has already been observed with Spitzer \citep{monaghan_2025}, yielding a dayside brightness temperature of $1862\pm60$ K. The existing observations favour either a highly reflective surface or the presence of a thin atmosphere, but do not yet discriminate between the two scenarios. If an atmosphere is present, reflective clouds could also contribute to a high apparent albedo and may be difficult to distinguish from surface reflection. However, high reflectivity from a bare lava surface alone is generally considered unlikely under such strongly irradiated conditions \citep{essack_2020}. Within our density-consistent simulations in Figure \ref{fig:regimes}, the vast majority of cases converge towards a solidified interior with no substantial atmosphere (Row B-C), with only a small fraction of solutions falling within the molten category, making this the target in our sample most consistently suggested to be in an advanced stage of crystallisation.

Examining the individual parameter dependencies, the escape efficiency (Panel C1) again sets a clear boundary: at $\eta = 10^{-2}$, the atmosphere is entirely stripped across all surviving cases. For the hydrogen inventory (Panel C2), an intermediate reservoir of 10~\hoceans\ yields the highest number of molten solutions, while both lower and higher inventories produce predominantly solid outcomes. This behaviour reflects the limited size of the mantle reservoir: independently of the nominal bulk hydrogen inventory, a comparatively thin mantle can only outgas and sustain a correspondingly modest atmosphere, which is then stripped rapidly, driving the solution towards the solid regime even at high nominal inventories. With respect to oxygen fugacity (Panel C3), reduced conditions ($\Delta$IW $< 1$) result exclusively in solid solutions, whereas more oxidised conditions ($\Delta$IW $\geq 1$) admit a small number of molten cases, though the solid category remains dominant even there. Across all explored values of the C/H ratios (Panel C4), solutions fall almost entirely within the solid regime, with only marginal exceptions at the highest C/H values. Finally, TOI-431 b is only consistent with the observed mass, radius, and age when modelled with a CRF $\geq 0.7-0.8$ R$_{\mathrm{p}}$ (Panel C5), in agreement with mass-radius relations that place this planet in a denser region of compositional space relative to an Earth-like interior, suggesting a significantly iron-enriched bulk composition.

\subsubsection{TOI-2260 b}

\begin{figure*}[t!]
    \centering
    \includegraphics[width=\linewidth]{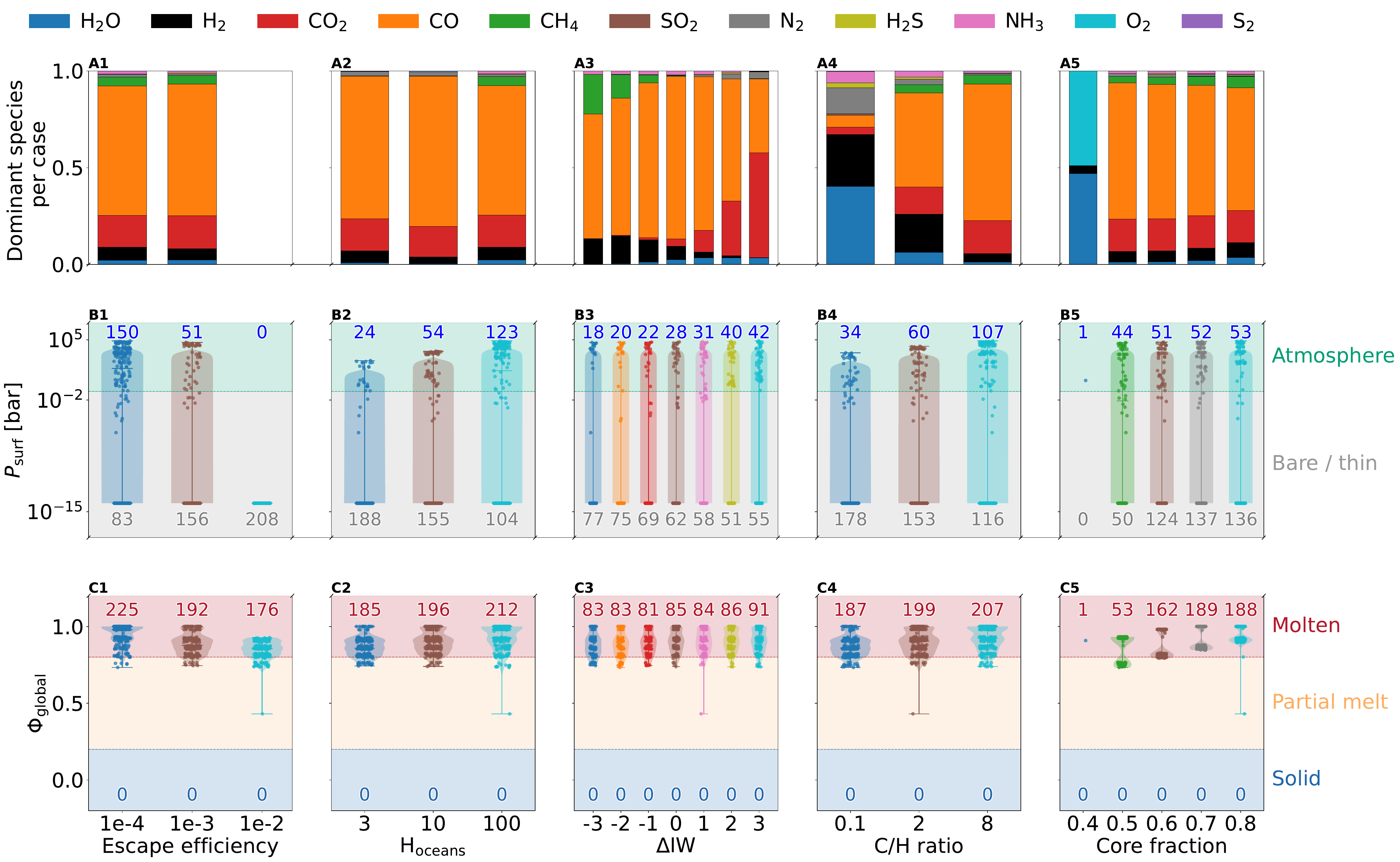}
    \caption{Summary of theoretical outcomes for TOI-2260 b across the 648 density-consistent cases, as a function of five input parameters: escape efficiency, initial water content \hoceans, oxidation state \fotwo, C/H ratio, and core fraction.}
    \label{toi2260}
\end{figure*}
Figure~\ref{toi2260} shows the distribution among drivers of density-consistent simulations for TOI-2260~b. We reiterate that the mass adopted for this planet is not a radial-velocity measurement, but a model-predicted value from \citet{giacalone_2022} (Section~\ref{sec:targets}); no substantial RV dataset is available for this target. The correspondingly large uncertainty on this predicted mass means that a substantial fraction of the grid simulations remain consistent with the adopted bulk density at the current age, and we caution that the results for TOI-2260 b should be interpreted in light of this additional layer of uncertainty relative to the other four targets. Despite this broad parameter degeneracy, the surviving solutions converge almost exclusively towards partially or fully molten interior states (Row C), making TOI-2260 b the target in our sample most potentially predicted to host a present-day magma ocean regardless of the assumed initial conditions. However, the majority of these partially molten solutions do not retain a substantial atmosphere (Row B), suggesting that volatile loss might dominate over outgassing for most of the explored parameter space, consistent with the young age of the system, during which enhanced stellar XUV irradiation drives efficient atmospheric escape that potentially outpaces volatile resupply from the molten interior.

Examining each parameter individually, the distributions are broadly uniform across most of the grid dimensions, with \fotwo (Panel A3) being the only driver that produces a clear trend in the median atmospheric composition, consistent with the behaviour previously described for TOI-1807 b. Escape efficiency (Panel C1) plays a decisive role in determining atmospheric retention: at $\eta = 10^{-2}$, virtually all solutions result in a completely stripped atmosphere, with the planet arriving at the stellar age  molten at that specific efficiency. As discussed above, a tenuous rock-vapour atmosphere may still persist in these cases, but such species are not included in our atmospheric model. In terms of interior structure (Panel C5), the density-consistent cases preferentially select core radius fractions above 0.5, and within this subset the interior is dominantly in the molten state.

\subsubsection{TOI-6255 b}
\begin{figure*}[t!]
    \centering
    \includegraphics[width=\linewidth]{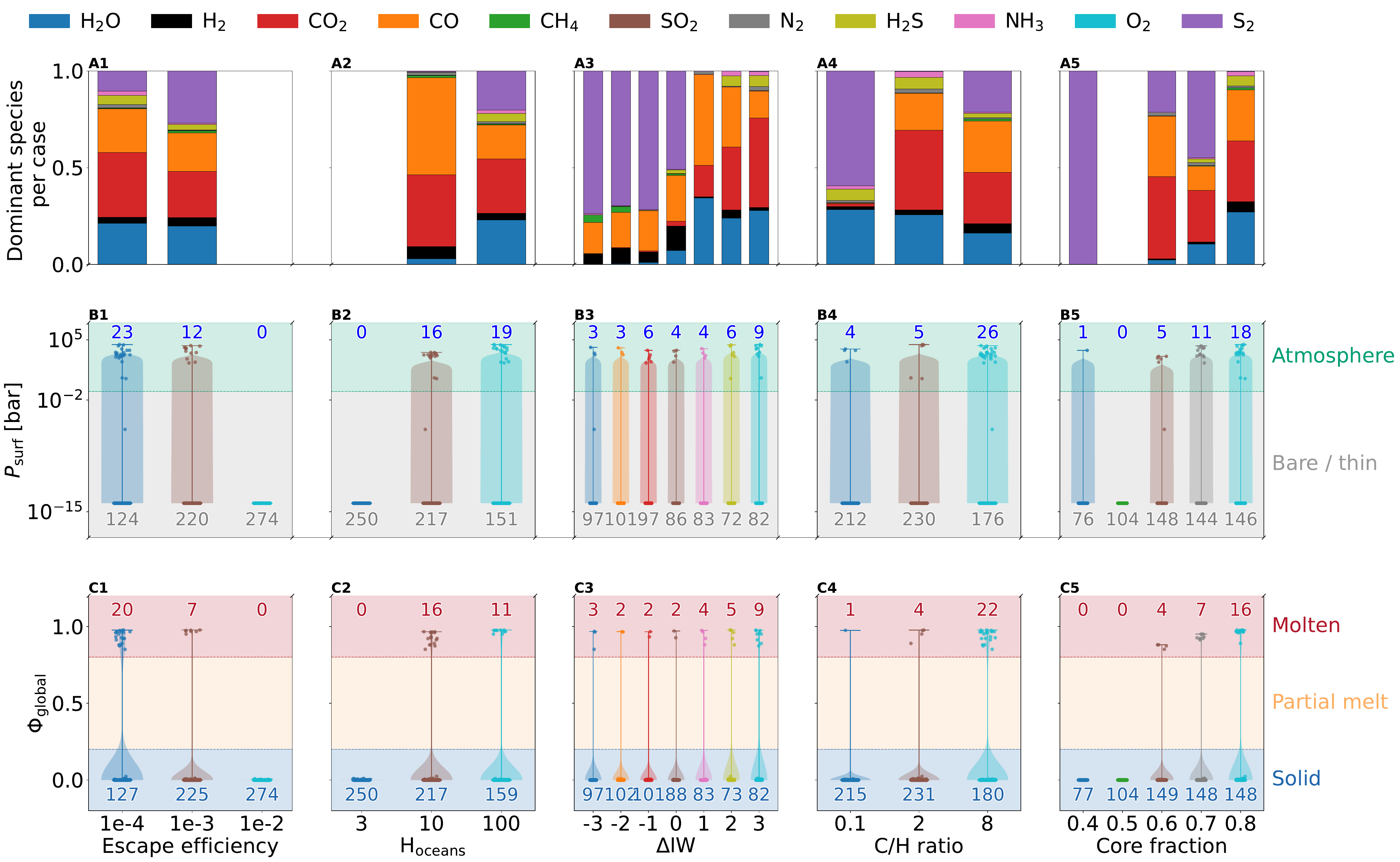}
    \caption{Summary of theoretical outcomes for TOI-6255 b across the 653 density-consistent cases, as a function of five input parameters: escape efficiency, initial water content \hoceans, oxidation state \fotwo, C/H ratio, and core fraction.}
    \label{toi6255}
\end{figure*}
TOI-6255 b is the smallest planet in the sample and also the one with the most tightly constrained mass and radius within the grid; Figure \ref{toi6255} shows the analysis of the 653 simulations that pass the density and age constraints, making it the target with the largest number of density-consistent cases. The overall distribution of interior states shows a strong preference for a solidified interior (Row C). Examining the individual parameter dependencies, the escape efficiency (Panel C1) follows the same trend seen across most of the targets: $\eta$ defines a clear threshold separating atmosphere-retaining from stripped solutions, with high escape efficiencies depleting the volatile inventory and driving the planet towards an atmosphere-free solid state. For the hydrogen inventory (Panel C2), the solid interior regime dominates across all explored values, with the highest number of solutions concentrated at intermediate volatile inventories and only 16 cases reaching a fully molten state. Oxygen fugacity (Panel C3) shows broadly similar distributions across the explored range, with no strong differentiation in interior state as a function of $\Delta$IW. For the C/H ratio (Panel C4), the majority of solid solutions cluster at intermediate values of $\sim$2, while the small number of molten cases is preferentially associated with higher C/H ratios. Finally, TOI-6255 b admits solutions across a core radius fraction range of $0.4$-$0.8\,R_\mathrm{P}$; the smallest core cases (CRF $= 0.4$ and $0.5$) yield exclusively solid solutions (Panel C5), and the number of molten cases increases progressively towards larger core radius fractions. Importantly, S$_{2}$ denotes as a dominant component of the atmosphere at reduced conditions and small to intermediate core sizes.

\subsubsection{TOI-2431 b}

\begin{figure*}[t!]
    \centering
    \includegraphics[width=\linewidth]{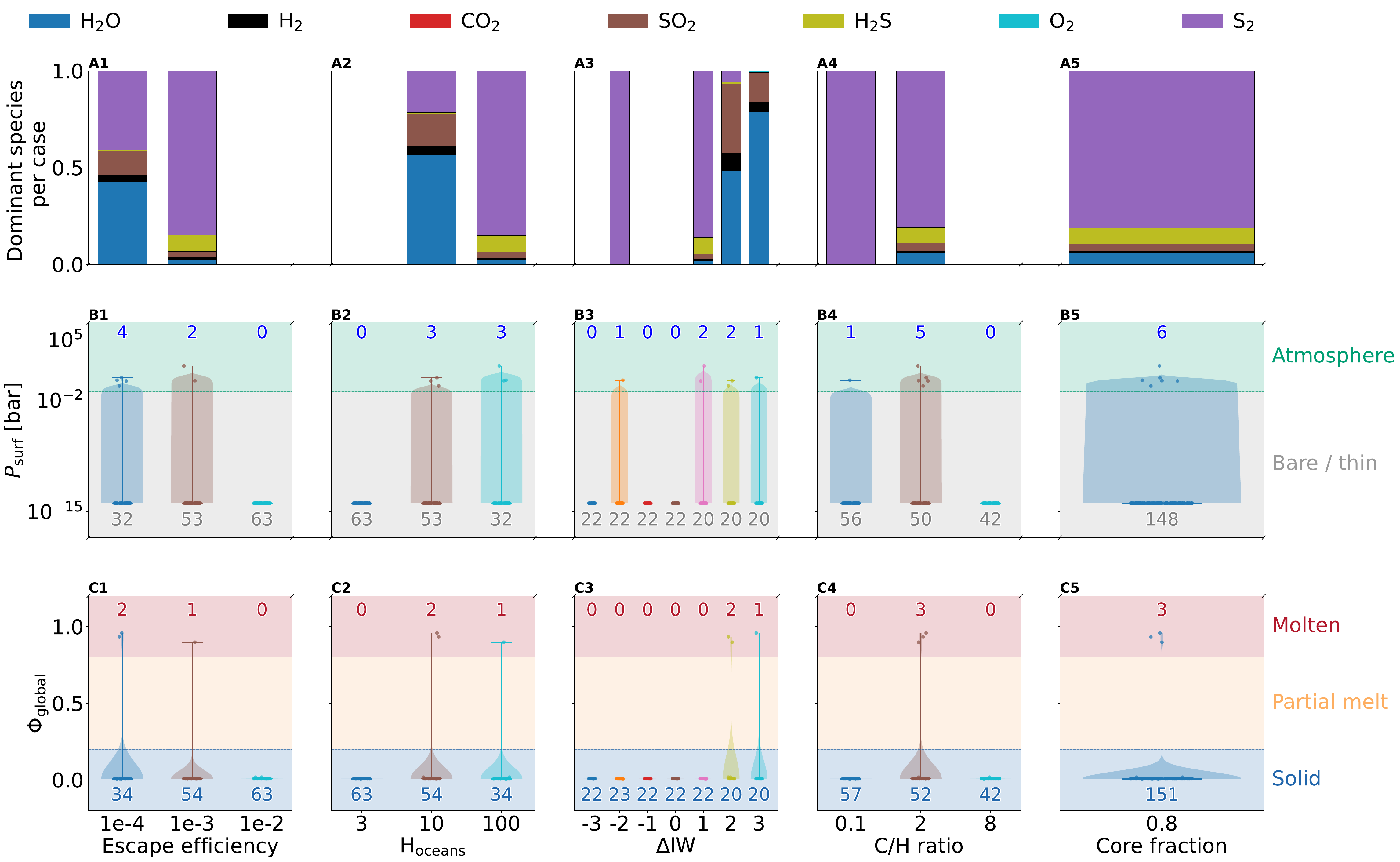}
    \caption{Summary of theoretical outcomes for TOI-2431 b across the 154 density-consistent cases, as a function of five input parameters: escape efficiency, initial water content \hoceans, oxidation state \fotwo, C/H ratio, and core fraction.}
    \label{toi2431}
\end{figure*}
TOI-2431 b is the most massive planet in the sample, and only 154 simulations achieve the density and age constraints, the smallest count among the five targets, see Figure \ref{toi2431}. This is primarily a consequence of the boundaries of the equation of state employed: modelling this planet pushes into extreme physical regimes where numerical instabilities become more frequent, reducing the fraction of grid points that yield converged solutions. These instabilities arise predominantly from the pressure-temperature conditions at the base of the mantle, which are independent of the presence or thickness of an atmosphere. As such, there is no evidence that the unconverged cases preferentially correspond to molten interiors, nor that the retained sample is systematically biased toward the solid regime. Despite this limitation, the density-consistent and converged cases are sufficient to identify clear trends across the parameter space, and the overall distribution strongly favours a solidified interior with no substantial atmosphere (Rows B-C).

Examining the individual parameter dependencies, the escape efficiency (Panel C1) follows the same behaviour seen across all targets, with $\eta$= 1e-2 acting as the threshold separating atmosphere-retaining from stripped solutions. For the hydrogen inventory (Panel C2), the maximum number of converged solutions occurs at intermediate volatile inventories, consistent with the pattern observed for the other older targets in the sample. Oxygen fugacity (Panel C3) shows negligible differentiation across the explored range, with the interior state distribution remaining largely insensitive to $\Delta$IW. The C/H ratio  (Panel C4) yields the highest number of surviving cases at the lowest explored ratio, with solution counts declining towards higher C/H values. Finally, TOI-2431 b is the most structurally constrained target in the sample (Panel C5): the density and age criteria are only satisfied at a core radius fraction of $0.8\,R_\mathrm{P}$, strongly suggesting a large-iron core, placing this planet in the high-density tail of the rocky planet population.

\subsection{Do They Retain an Atmosphere?}
\label{sec:results_atm}

\begin{figure}[h]
    \centering
    \includegraphics[width=\linewidth]{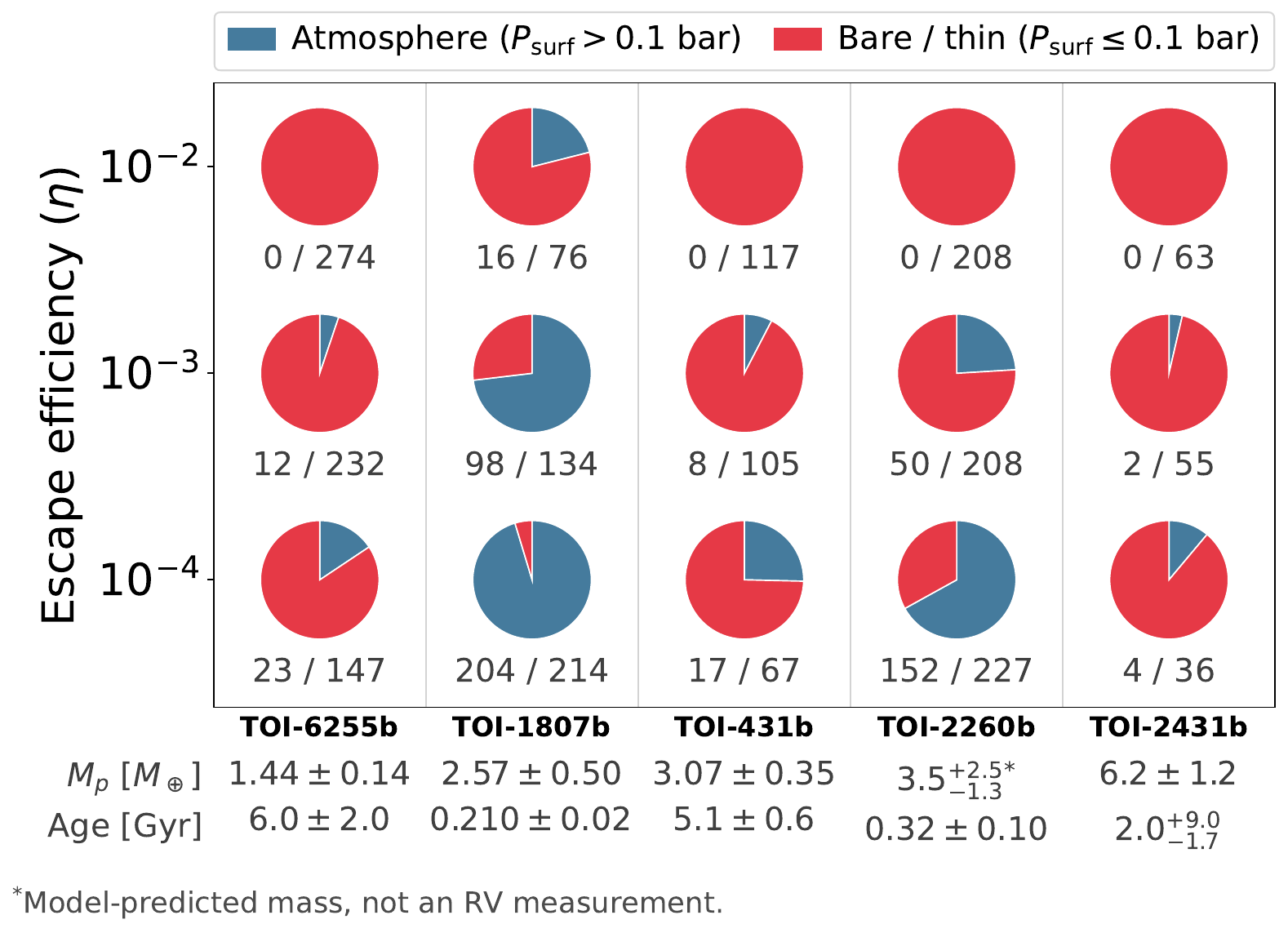}
    \caption{Escape efficiency $\eta$ of the density-consistent cases for each target planet, ordered by increasing planetary mass from left to right. Each pie chart corresponds to one value of $\eta$ for one planet and shows the fraction of cases that retain a substantial atmosphere ($P_\mathrm{surf} > 0.1$\,bar, blue) and the fraction that end as bare rocks or with a thin atmosphere ($P_\mathrm{surf} \leq 0.1$\,bar, red). The numbers below each pie give the number of atmosphere-bearing cases over the total number of density-consistent cases for that combination. The planetary mass and system age are listed below each planet name.}
    \label{fig:escape}
\end{figure}

To tackle this question we identified the minimum threshold of escape efficiency required to retain a substantial atmosphere. Figure~\ref{fig:escape} shows the distribution of density-consistent simulations across the explored $\eta$ values, colour-coded by whether the simulation produces an atmosphere-bearing ($P_\mathrm{surf} > 0.1$~bar) or a bare to thin ($P_\mathrm{surf} \leq 0.1$~bar) outcome. Across all planets but TOI 1807b, $\eta = 10^{-2}$ is occupied exclusively by bare/thin cases, establishing this value as a suggestive threshold above which atmospheric stripping is complete regardless of the volatile inventory or interior state in the energy-limited escape but sensitive on very young systems as the one described for TOI 1807 b \citep{barber_2025}. At the lowest efficiency explored, $\eta = 10^{-4}$, atmosphere-retaining solutions become more prevalent for the two young targets. For the older targets TOI-431 b, TOI-2431 b and TOI-6255 b, atmosphere-bearing cases are considerably rarer even at the lowest efficiencies, with bare rock solutions dominating at $\eta > 10^{-3}$ and only a small fraction of atmosphere-retaining cases appearing at $\eta = 10^{-4}$, reflecting the reduced volatile budget available after longer periods of atmospheric erosion and interior cooling.

\begin{figure}[h]
    \centering
    \includegraphics[width=\linewidth]{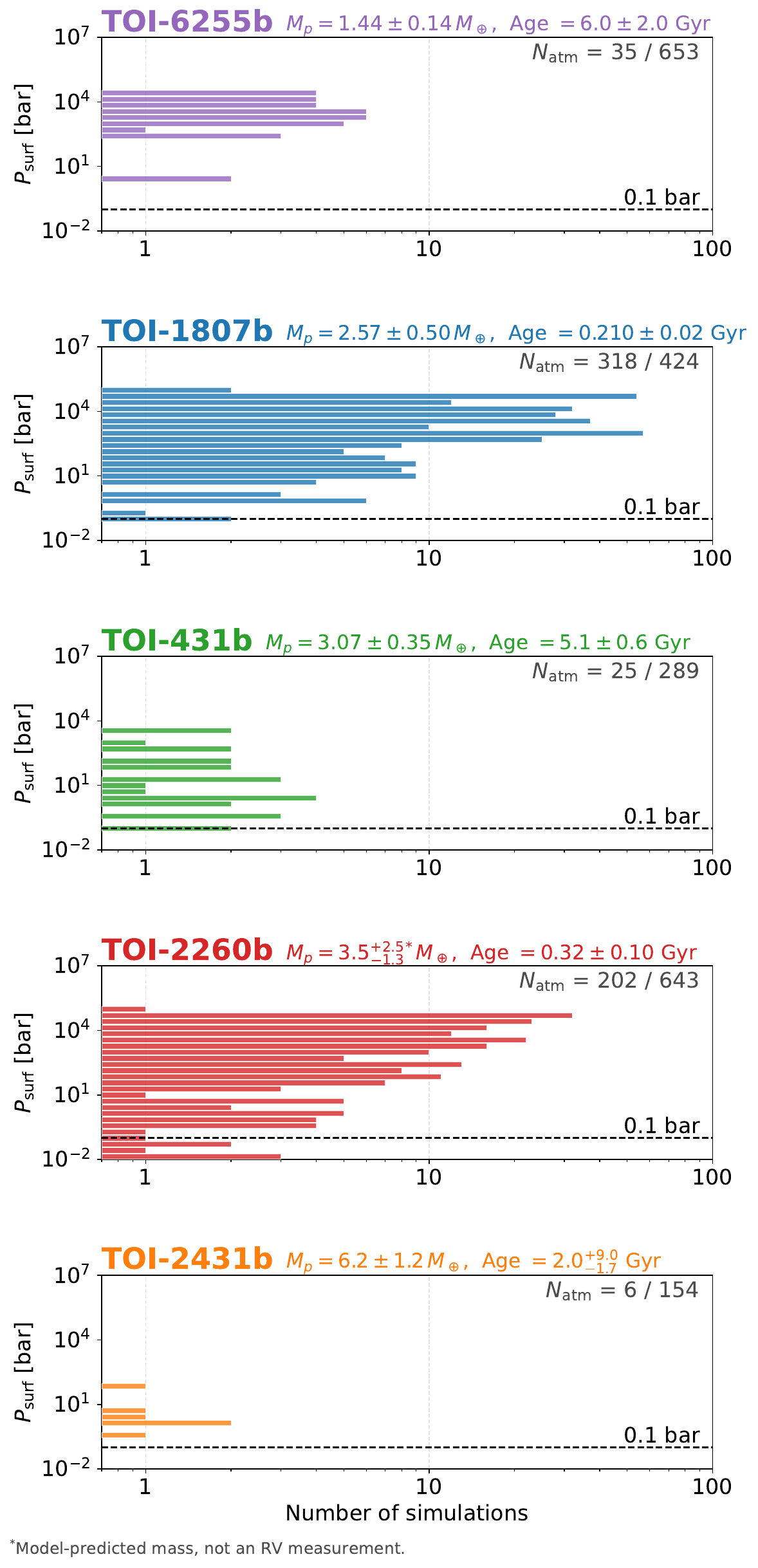}
    \caption{Surface pressure distributions of the atmosphere-retaining cases ($P_\mathrm{surf} > 0.1$\,bar) among all density-consistent simulations for each target planet. Each horizontal bar represents a $P_\mathrm{surf}$ bin, with bar length indicating the number of simulations. The dashed line marks the $P_\mathrm{surf} = 0.1$\,bar threshold for reference. The annotation $N_\mathrm{atm}$ gives the number of atmosphere-retaining cases out of the total density-matching cases. The distributions illustrate the diversity in predicted atmospheric mass across targets, highlighting which planets favour thick versus tenuous atmospheres among the cases that retain one.}
    \label{fig:psurf}
\end{figure}

Similarly, Figure~\ref{fig:psurf} shows the distribution of surface pressure across all density-consistent simulations for each target, restricted to cases in which an atmosphere is retained ($P_\mathrm{surf} > 0.1$~bar). The two young targets, TOI-1807 b and TOI-2260 b, host the largest atmospheric fractions in the sample ($N_\mathrm{atm} = 318/424$ and $202/643$, respectively). For TOI-1807 b, the pressure distribution is broad, spanning from $\sim 10^{-1}$ to $10^5$~bar, with a pronounced concentration of cases between $10^3$ and $10^5$~bar, indicative of thick, volatile-rich 
envelopes. TOI-2260 b follows a similar trend, though its distribution extends to lower pressures and includes a small fraction of cases falling below the 0.1~bar threshold, suggesting that a subset of solutions sits at the boundary between a tenuous and a substantial atmosphere.

Among the older targets, TOI-6255 b retains an atmosphere in 35 out of 653 density-consistent simulations. Its pressure distribution spans from $\sim 10^0$ to $10^4$~bar, with a clear peak at $\sim 10^3$~bar, indicating that when an atmosphere is present it tends to be moderately to highly pressurised. TOI-431 b shows 25 atmosphere-bearing cases out of 289, with a comparatively flat distribution extending up to $\sim 10^3$~bar and a concentration of cases near $10^0$--$10^1$~bar, consistent with thinner outgassed envelopes. TOI-2431 b has the fewest atmosphere-retaining solutions in the sample, 
with only 6 out of 154 density-consistent simulations exceeding the 0.1~bar threshold; in all of these cases the surface pressure is clustered around $\sim 10^0$~bar, pointing to a very limited volatile budget in the converged solutions.

\subsection{Atmospheric Composition}
\label{sec:results_comp}

\begin{figure}[h]
    \centering
    \includegraphics[width=\linewidth]{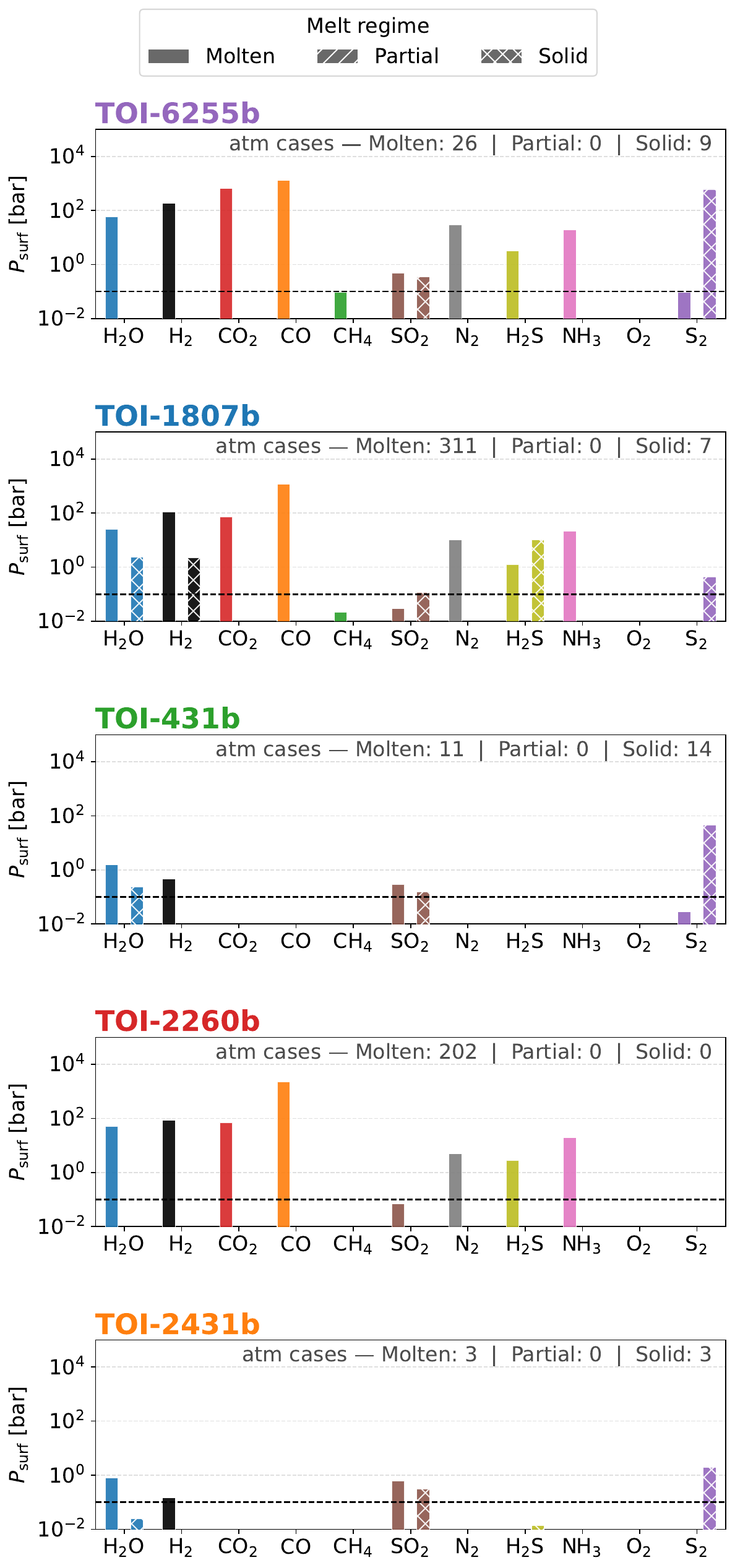}
    \caption{Median partial pressure of each atmospheric species per melt regime (molten: solid bars; partial melt: hatched bars; solid: cross-hatched bars) for the atmosphere-retaining cases ($P_\mathrm{surf} > 0.1$\,bar) of each target planet, ordered by increasing planetary mass. The number of cases in each melt regime is indicated in the panel title. The dashed horizontal line marks $P_\mathrm{surf} = 0.1$\,bar for reference. }
    \label{fig:species_by_regime}
\end{figure}

When examining the atmospheric composition across the different interior melt regimes, Figure~\ref{fig:species_by_regime} shows the median partial pressure of each atmospheric species for all atmosphere-bearing cases ($P_\mathrm{surf} > 0.1$~bar) across the five targets. TOI-6255 b and TOI-1807 b exhibit the richest atmospheric compositions, with median partial pressures spanning up to $\sim 10^4$~bar across multiple species. In both planets, \water, H$_2$, \cotwo, and CO are the dominant constituents, followed by secondary contributions from N$_2$, H$_2$S, NH$_3$, and S$_2$. For nearly all species, the median partial pressure is systematically higher in the molten regime than in the solid one. This does not reflect a larger cumulative outgassed inventory, since crystallisation expels dissolved volatiles and a solidified mantle has released most of its budget. Instead, it reflects a feedback between volatile retention and the melt state of the mantle. A thick atmosphere provides the greenhouse blanketing that keeps the surface hot and the mantle molten. In turn, the molten mantle dissolves part of the volatile inventory and, being in chemical equilibrium with the atmosphere, resupplies volatiles as they are lost to escape. Once the interior solidifies, this reservoir is no longer available, escape strips an atmosphere that can no longer be replenished, and the loss of blanketing further favours cooling. The molten regime is therefore both a consequence and a cause of retaining a substantial atmosphere.~\citep{bower_retention_2022, nicholls_magma_2024,nicholls_thesis_2026}. A notable exception is S$_2$, which shows a partial pressure larger by roughly
an order of magnitude in the solid regime than in the molten one. Because our
atmospheres are isochemical, with composition set by outgassing and no
chemistry computed in the atmosphere itself, this reflects a shift in the
outgassed sulphur speciation as the interior cools and crystallises, rather
than gas-phase thermochemistry~\citep{panagiotou_2026}.  We interpret this as arising from a combination of dissolution and escape effects. S-bearing species remain preferentially dissolved in the melt and are only outgassed significantly during the later stages of crystallisation, in contrast to \cotwo, CO, and H$_2$, which are released earlier while the interior is still substantially molten. Once outgassed, S$_2$ is also considerably heavier (by a factor of $\sim$1.5 relative to \cotwo, and $>3$ relative to CH$_4$, \water, and NH$_3$), making them more resistant to atmospheric escape than these lighter, earlier-outgassed species. The combination of late-stage outgassing and a higher molecular weight therefore favours the relative retention of S$_2$ in the atmosphere, consistent with the shift towards S$_2$-dominated compositions in the solid regime. This suggests that S$_2$ may be used as an atmospheric tracer of interior melt state. We note, however, that this result is obtained under a fixed, Earth-like S/H mass ratio (Section~\ref{tab:params}), such that the absolute sulphur budget scales only with the hydrogen inventory.

Notably, S$_2$ is absent from the TOI-2260 b panel, consistent with the absence of solid-interior solutions for this planet and supporting the interpretation that S$_2$ enrichment is linked to the solid end-member state. CO reaches the highest median partial pressures among all species for this target, exceeding $10^3$~bar in both the molten regime.

TOI-431 b and TOI-2431 b present a markedly different picture. Both planets show only a handful of atmosphere-bearing cases, and the median partial pressures of most species fall at or below the 0.1~bar threshold. The only species reaching appreciable levels are SO$_2$ and S$_2$, the latter particularly in the solid regime of TOI-431 b, while \water, H$_2$, \cotwo, and CO remain at very low pressures. This overall picture is consistent with the bare rock or thin atmosphere scenario inferred for both targets from their melt state distributions.

We note that these are 1D models, and the surface pressures reported here correspond to an atmosphere resulting from the globally averaged instellation flux, rather than conditions at a specific location on the planet's surface (e.g. the sub-stellar point or terminator). As such, they do not resolve the strong day--night pressure and temperature contrasts predicted by atmospheric circulation models of synchronously rotating lava planets \citep{kite_2016, nguyen_2020}, which find that atmospheres can be largely confined to the dayside, with negligible pressure on the nightside. Furthermore, the evolutionary phase considered here corresponds to the early magma-ocean evolution, when the assumption of a globally averaged instellation is expected to be more representative than in the later synchronously rotating state. Consequently, the reported surface pressures should be interpreted as global mean values rather than local atmospheric conditions.

\subsection{Predicted Observables and Detectability}
\label{sec:results_detect}
\begin{figure}[h]
    \centering
    \includegraphics[width=\linewidth]{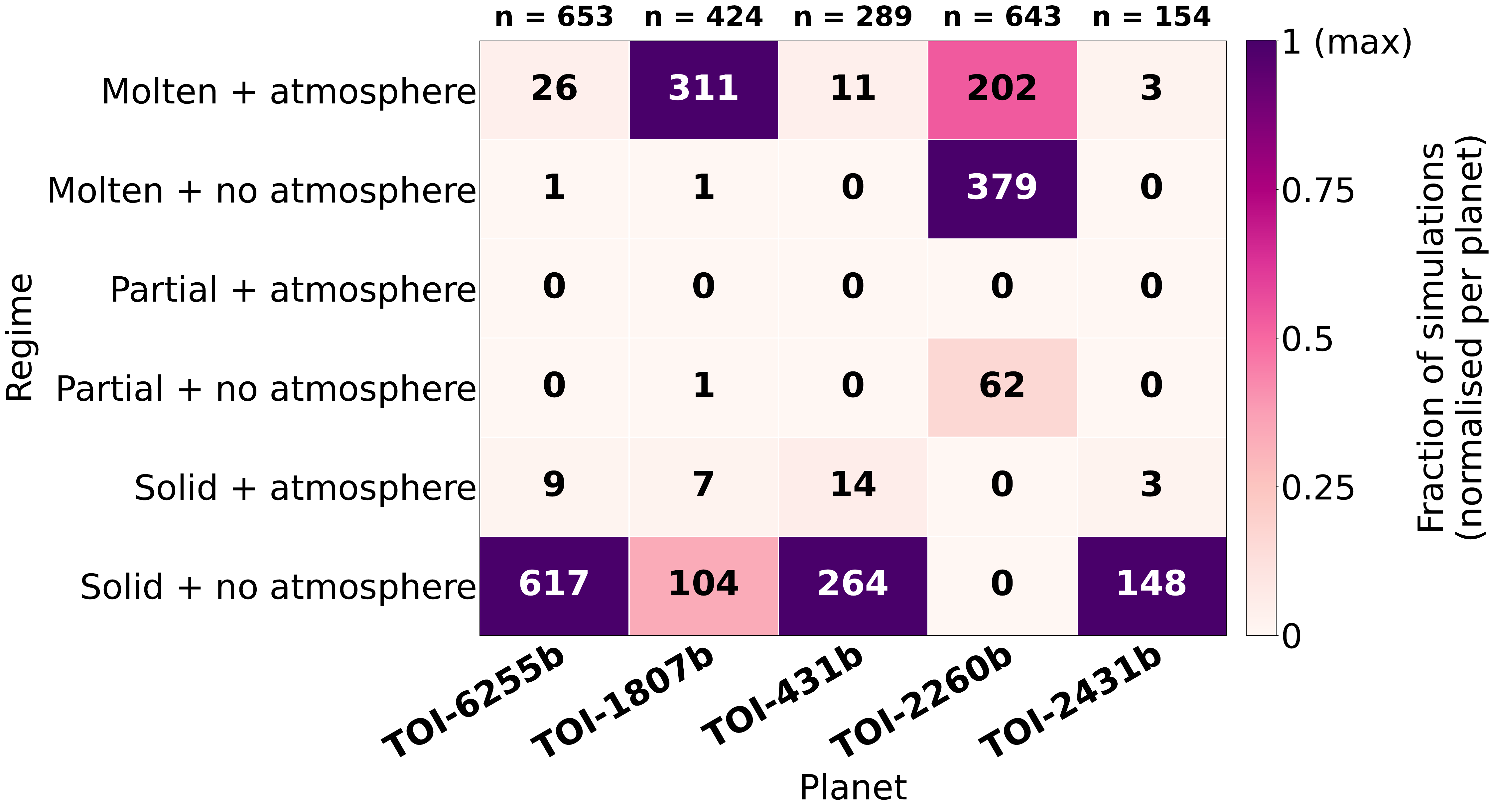}
    \caption{Predicted detectability metric: Interior-atmosphere regime summary for all density-consistent cases across the five target planets, ordered by increasing planetary mass. Each cell shows the number of simulations falling into each interior-atmosphere regime combination (molten, mushy, or solid interior; with or without a substantial global atmosphere), normalised per planet by the maximum count across all regime combinations. The total number of surviving cases per planet ($n$) is indicated on top of the plot.}
    \label{fig:regime_interior}
\end{figure}

Once the parameter-by-parameter sensitivity analysis was completed for each target, we summarised the model outcomes as the frequency with which density-consistent simulations fall into each interior–atmosphere regime, in order to translate our results into predictions for potential observables. These frequencies reflect how extensively each outcome is represented within the explored parameter space, not formal probabilities, as they depend on the sampling of the grid (Section~\ref{sec:methods}). Fig.~\ref{fig:regime_interior} presents this distribution across all five planets simultaneously. The panel shows the fraction of density-consistent simulations falling within each of the six interior-atmosphere categories defined in Sect.~\ref{sec:filtering}, normalised per planet, providing a visual indication of which thermal state is preferred for each target given the explored parameter space, with S$_2$ standing out as the species most closely tracing a solid-state interior. Together with the results from Figure \ref{fig:species_by_regime}, this provides a compact summary of the predicted thermal states and the atmospheric constituents most likely to be accessible to MIRI-LRS for each of the five targets.

The predicted thermal state has direct implications for the expected MIRI-LRS phase curve. For the targets where a solid, atmosphere-free interior dominates the final outcomes, the absence of a substantial atmosphere to redistribute heat implies a very cold nightside and a correspondingly large day-night temperature contrast, which should manifest as a large-amplitude phase curve. Conversely, for targets and regimes where a molten interior with a significant atmosphere is favoured, more efficient heat redistribution is expected to suppress this contrast, resulting in a comparatively small phase curve amplitude. The regime probabilities summarised in Fig.~\ref{fig:regime_interior} therefore translate directly into a testable prediction for the phase curve amplitude expected from each target.

\section{Discussion}
\label{sec:discussion}

\subsection{Trends Across the Planet Sample}
\label{sec:discussion_trends}
Taken together, the results across the five targets reveal a clear dichotomy between the two younger planets and the three older ones. TOI-1807 b and TOI-2260 b are the only targets that show >200 simulations retaining an atmosphere and a molten interior simultaneously, with atmospheric masses that are significant across a broad range of volatile inventories. These two planets also display an abundant presence of \water\ in the atmosphere, reflecting the combined effect of active outgassing from a still-molten mantle \citep{bower_retention_2022, piaulet_2025, lichtenberg_super-earths_2025, krissansen-totton_predictions_2022} and the short time available for atmospheric escape and interior solidification relative to their age \citep{hamano_2015,nicholls_magma_2024}.

In contrast, the three older targets converge predominantly towards bare-rock or thin-atmosphere solutions. This is consistent with the expectation that, given sufficient time to cool, the younger targets in our sample would likewise evolve towards the solidification. TOI-431 b and TOI-2431 b are particularly noteworthy in that they only admit numerical solutions with large core radius fractions ($\geq 0.7\,R_\mathrm{P}$ and $0.8\,R_\mathrm{P}$, respectively), a structural constraint that is physically consistent with their atmospheric properties: a large core implies a small silicate mantle and therefore a limited melt reservoir available for volatile outgassing \citep{stixrude_2014,nicholls_magma_2024,shorttle_2024,Suer2023FrEaS, boley_fizzy_2023,salvador_magma_2023, lichtenberg_super-earths_2025}. For these two planets, the oxygen fugacity (Panel C3 in Figures \ref{fig:regimes} and \ref{toi2431}) also plays a more discernible role than for the rest of the sample, oxidized conditions (\fotwo $\geq$IW) tend to include solutions in the molten regime, while reduced conditions produce fully solidified interiors, with a corresponding imprint on the atmospheric composition \citep{bower_retention_2022, Boer2025ApJ, nicholls_magma_2024, van_Dijk_2026, sastre_2026}. These results provide empirical constraints on the interior structure of ultra-short-period rocky exoplanets, and in particular suggest that the subset of targets requiring large core radius fractions to satisfy the observed mass-radius relation may be systematically poor in outgassed volatiles and unlikely to host detectable atmospheres \citep{krissansen-totton_predictions_2022}.

As discussed in Section \ref{sec:results_atm}, four of the five targets consistently lose their atmosphere ($P_\mathrm{surf} \lesssim 10^{-12}$~bar) once the escape efficiency reaches $\eta = 10^{-2}$, regardless of the volatile inventory, interior melt state, or oxidation conditions. This threshold therefore acts as a hard boundary in parameter space: no atmosphere-bearing solution survives at this efficiency across any of the planets in the sample with ages $\geq$320 Myr, suggesting that energy-limited escape at this rate and time is sufficient to strip the entire volatile budget over the system lifetime in agreement with previous studies that showed similar trends with larger efficiencies in different compositions \citep{bello_2025, owen_2019, luger_extreme_2015}. This result is robust across the full range of planetary masses, ages, and interior structures explored here, suggesting that $\eta = 10^{-2}$ represents an upper limit on the escape efficiency compatible with the retention of a detectable atmosphere for ultra-short-period rocky planets in this irradiation regime. Although lower escape efficiencies are often treated as a limiting case, the energy-limited regime predicts that heavier atmospheres lose mass less efficiently, since removing a given amount of material requires more energy as the mean molecular weight increases \citep{wordsworth_redox_2018, yoshida_2024, chaterjee_2026}. Additionally, radiative line cooling, molecular emission, and ionization losses can further remove a substantial fraction of the energy deposited by XUV irradiation, lowering the atmospheric temperature and thereby reducing escape rates \citep{nakayama_2022, yoshida_2024, chaterjee_2026}. Because the atmospheres modelled here are dominated by volatiles outgassed from a magma ocean rather than light primordial H/He gas, they can attain high mean molecular weights. In addition, escape efficiency is known to depend on planetary properties and atmospheric composition \citep{rogers_2021} motivating the lower values explored here ($\eta \lesssim 10^{-3}$-$10^{-4})$. Atmospheres can therefore persist even at high irradiations \citep{ito_2021, nakayama_2022}.

\subsection{Geophysical implications for the target planets}
\label{sec:implications}

The results presented here represent a step toward constraining the geophysical state of USP exoplanets. A key outcome is that the persistence of a detectable atmosphere is closely tied to the melt state of the interior, which itself depends on planetary age, mass, composition and tectonic regime \citep{stixrude_2014, nicholls_magma_2024, foley_tectonic_2018}. Linking  thermal emission observations to the underlying geophysical processes governing planetary evolution. 

Furthermore, for bare-rock or weakly outgassing USP planets, extreme stellar irradiation introduces an additional layer of geodynamic and atmospheric complexity. On the dayside of these planets, surface temperatures can be high enough to directly vaporize silicate rock, producing a mineral-vapour atmosphere. Supersonic winds driven by the resulting day-night pressure gradient are then expected to advect this rock vapour towards the nightside \citep{castan_2011,kite_2016, nguyen_2020,nguyen_2025}, where it may recondense \citep{zilinskas_2022, meier_2023}, further complicating the interpretation of any thin or asymmetric atmosphere inferred for these targets. In addition, the strong temperature contrast can drive hemispheric asymmetries in mantle convection \citep{meier_hemispheric_2021,meier_2023,Meier2024JGRE12908491M}, potentially leading to spatial variations in volcanic activity and volatile transport. As a result, outgassing may become concentrated on the dayside or nightside, producing atmospheres with distinct chemical compositions between hemispheres, on top of the rock-vapour dynamics described above. Such asymmetries could leave observable signatures in phase-curve offsets and secondary-eclipse spectra, complicating their interpretation while simultaneously offering a unique window into the interior dynamics of these planets.

The oxygen fugacity (\fotwo) values recovered from our model grid provide an additional geophysical diagnostic. The strong sensitivity of both the interior melt state and atmospheric composition to \fotwo\  particularly evident for TOI-431 b and TOI-2431 b, suggests that future atmospheric observations may offer a means of constraining the redox state of these planets. If atmospheric chemistry can be characterized well enough to distinguish between oxidized and reduced scenarios, it may become possible to infer the redox conditions of the underlying melt and, consequently, gain insights into the mineralogical nature of the source material.

\subsection{What Can JWST Realistically Constrain?}
\label{sec:discussion_jwst}

Program 8864 \citep{dang_proposal_2025} is a comparative thermal emission full-orbit phase curve study, providing access to both thermal and energy budget properties as well as atmospheric dynamics and circulation for all targets. This builds on a growing set of comparable JWST phase-curve programs targeting individual USP and lava-planet targets, including TOI-731 b \citep{zhang_jwst_2021}, TOI-561 b \citep{teske_jwst_2023}, and K2-141 b \citep{espinoza_jwst_2021, dang_2021}. It also complements two larger JWST survey programs targeting the broader population of small, highly irradiated rocky worlds: the Hot Rocks Survey \citep{diamond_jwst_2023}, which tests nine irradiated terrestrial exoplanets for the presence of atmospheres via secondary eclipse photometry, and the COMPASS survey \citep{batalha_jwst_2021}, which targets the transmission spectra of 11 small (1--3~$R_\oplus$) planets to link atmospheric composition with planetary demographics. In the context of the interior regimes defined in this work, these observations can be interpreted following a natural diagnostic sequence that progressively narrows down the possible interior and atmospheric states of each planet.

The first and most powerful diagnostics are the dayside and nightside temperatures. A dayside temperature lower than expected may indicate the presence of an atmosphere, due to heat redistribution \citep{mansfield_2019, koll_2019}. Cooler-than-expected dayside temperatures have now been reported for previously studied lava worlds \citep{teske_thick_2025}, making a similar signature a plausible outcome for the targets in our program. Conversely, a comparatively warm nightside is generally interpreted as evidence for atmospheric heat redistribution \citep{hammond_2025}. However, at the high temperatures considered here, heat redistribution is not expected to be fully efficient, as observed for hot and ultra-hot Jupiters, so even substantial atmospheres may produce only modest nightside warming \citep{splinter_2025}. Critically, a warm nightside is not an unambiguous atmospheric signature for these targets: a molten interior can itself sustain substantial nightside emission through internal heat transport, independent of any atmosphere \citep{meier_2023, meier_hemispheric_2021, boukare_2025}, introducing a degeneracy between atmospheric redistribution and an internally heated magma ocean as the origin of nightside warming. Distinguishing between the two scenarios therefore requires considering the dayside and nightside temperatures jointly: a warm nightside combined with a dayside temperature near the maximum expected for an airless, irradiated surface would favour an internally heated, molten interior over atmospheric heat redistribution.

If an atmosphere is present, the dayside thermal emission spectrum becomes a key diagnostic, potentially revealing molecular absorption or emission features indicative of atmospheric composition (e.g., SO\textsubscript{2} at 7-10~\textmu m). If no atmosphere is detected, the thermal emission still constrains the surface albedo, offering insight into the nature of the underlying surface \citep{guimond_mineralogical_2023, hammond_2025}.

 The predicted atmospheric compositions shown in Figure~\ref{fig:species_by_regime} demonstrate that the mean partial pressures of different species vary systematically with interior melt state, providing a direct link between the observed emission features and the underlying interior regime. In particular, the contrast in mean atmospheric composition as a function of input parameters is most pronounced for the oxidation state \fotwo and escape efficiency, especially for planets with large core fractions, where differences across parameter bins are most clearly distinguishable. Core radius fraction itself is largely degenerate in terms of atmospheric composition, as the predicted mean compositions 
across different core fraction bins are similar for most targets. The C/H ratio shows a strong compositional gradient across parameter bins, but primarily for the younger planets in the sample; for the remaining targets the spread around the mean is too large to draw robust conclusions. Similarly, \hoceans shows limited discriminating power for most targets.

In summary, the phase curve observations from Program 8864 \citep{dang_proposal_2025} suggest a natural, sequential approach to constraining the interior state of each target:
\begin{enumerate}
    \item Nightside temperature: determine whether the planet is in a molten or solid/partial melt state, and whether an atmosphere is present.
    \item Dayside temperature and emission: if an atmosphere is detected, constrain its bulk composition and rule out atmosphere-free regimes. The phase curve morphology may also provide qualitative evidence for atmospheric heat redistribution through features such as hotspot offsets.

    \item Compositional diagnostics: use the inferred atmospheric composition to place constraints on the oxidation state and escape efficiency, particularly for planets with large cores, and on the C/H ratio for the youngest targets in the sample; the relative abundance of S$_{2}$ provides an additional discriminant between molten and non-molten interiors, given its systematic enhancement in the solid regime.
\end{enumerate}

This sequential framework highlights the strength of the multi-planet, comparative approach of Program 8864 \citep{dang_proposal_2025}: by observing targets spanning a range of masses, ages, and equilibrium temperatures, it becomes possible to disentangle parameter degeneracies that would be irreducible from a single-planet observation alone.
This comparative framework also provides a basis for prioritizing and designing future follow-up observations. Beyond predicting the thermal evolution of the five Program 8864 targets, our results can identify which planets are most promising for atmospheric characterization and which are better suited to surface-composition studies. The predicted interior regimes, surface pressures, and thermal emission can further guide the selection of observing modes and integration times, helping maximize the scientific return of future JWST and ground-based campaigns.

\subsection{Caveats and Model Limitations}
\label{sec:discussion_caveats}

The following outlines the main simplifying assumptions adopted in this work and their expected impact on the results. The thermal structure is modelled as a 1D average dayside profile. Day-night heat redistribution is therefore not captured, which is a reasonable first approximation given that these models focus on atmosphere-interior interactions rather than on phase-dependent thermal properties, and given that the primary observables from Program 8864 \citep{dang_proposal_2025} will probe thermal emission across the orbit and the secondary-eclipse spectrum, with the nightside spectrum accessible if the S/N permits. We therefore include day-night heat redistribution and a first-order treatment of tidal heating, adopting a modest constant heating rate corresponding to an orbital eccentricity of $\epsilon=0.05$, consistent with conditions under which sustained tidal dissipation in a mush layer may occur \citep{herath_2024}. While this simplified treatment does not capture the coupled evolution of tides, the interior, and the orbit, such effects are beyond the scope of the present study and have been explored in recent coupled models \citep{van_Dijk_2026, herath_2026}. We therefore regard our treatment as a conservative first-order approximation and expect more complete tidal-thermal-orbital coupling to primarily increase the range of molten solutions.

The volatile inventory is fixed at the start of each simulation, meaning that potential stellar abundances \citep{guimond_2024}, late accretion and post-formation volatile delivery are not explicitly modelled. However, by spanning a wide range of initial conditions across the model grid  including $H_\mathrm{oceans}$, C/H ratio, and $\Delta$IW  we effectively sample the diversity of outcomes that could arise from different accretion and delivery histories.

We note that rock vapour speciation, which is expected to contribute to the observable emission of hot, molten rocky planets \citep{zilinskas_2022, Piette2023, van_buchem_2025, Zilinskas_2025, van_buchem_2026}, and may also play a significant role in shaping atmospheric temperature structures and cooling timescales, is not included in the current framework, and represents a natural extension of this work. Additionally, volatile sequestration through melt trapping or crystal inclusions \citep{sim_volatile_2024, joshua_2024} is not modelled; such processes, if included, could reduce outgassing efficiency, potentially yielding thinner atmospheres and shorter magma ocean lifetimes \citep{lebrun_thermal_2013, salvador_relative_2017, nikolaou_what_2019, bower_linking_2019}.
Our atmospheres are isochemical: their composition is fixed by the outgassing equilibrium at the surface, and no chemistry is computed within the atmosphere. Outgassing is expected to dominate the bulk atmospheric composition
\citep{panagiotou_2026}, but gas-phase thermochemistry could still change the
abundances of individual species. In particular, S$_2$ is thermochemically
favoured at intermediate temperatures \citep{hobbs_2021}, so including atmospheric chemistry may
make the S$_2$ enhancement we find in the solid regime even more pronounced.

Finally, uncertainties in stellar age are not expected to significantly affect our results, as the relevant phases of magma ocean evolution occur on timescales much shorter than the age uncertainties for most of our systems (Table~\ref{tab:targets}). Despite these assumptions, the broad parameter space explored across the model grid ensures that the physical trends identified in this work including the dependence of atmospheric retention on escape efficiency, the compositional differences across oxidation states, and the distribution of interior regimes across targets remain robust to the model simplifications discussed above.

\section{Conclusions}
\label{sec:conclusions}

We performed a sensitivity analysis of magma ocean evolution pathways using the coupled interior--atmosphere framework \texttt{PROTEUS} for the five rocky exoplanet targets of JWST program 8864: TOI-1807 b, TOI-2260 b, TOI-431 b, TOI-6255 b, and TOI-2431 b. Using existing observational constraints on planetary mass, radius, age, semi-major axis, and stellar mass, we modeled the thermal and volatile evolution of each planet across a wide grid of initial conditions. Our main conclusions are as follows:

\begin{itemize}

    \item The bare rock scenario, with negligible volatile envelope and $\phi \leq 0.2$, dominates the surviving solutions for most targets. TOI-1807 b and TOI-2260 b are the notable exceptions: both exhibit a substantial fraction of simulations in a  globally fully or partially molten state with an atmosphere present, with TOI-2260 b showing all density-consistent solutions in the molten regime. These two planets also produce the most atmosphere-retaining cases overall, with H$_2$O and H$_2$ as the dominant species, and tend towards more massive atmospheric outcomes among atmosphere-bearing cases.

    \item TOI-431 b, TOI-2431 b, and TOI-6255 b are dominated by bare-rock solutions, with only a small number of density-consistent cases retaining a detectable atmosphere. For TOI-6255 b, the few density-consistent cases that do retain an atmosphere tend toward high surface pressures.

    \item Escape efficiency under the energy-limited regime acts as a fundamental parameter controlling whether a planet retains or loses its atmosphere. A clear threshold of $\eta \leq$10$^{-2}$ in escape efficiency separates atmosphere-retaining from bare rock outcomes across all  with ages $\geq$320 Myr. If residual atmospheres are found on the modelled super-Earths, this directly constrains escape physics.

    \item Among atmosphere-bearing cases, S$_2$ is exclusively produced in the solid interior regime across all targets, making its detection a potential diagnostic of a solidified mantle for an initial Earth-like S/H.

    \item Planets with large core fractions strongly favour bare rock outcomes. However, in the rare cases where an atmosphere is present, the larger geochemical contrast across input parameter bins makes atmospheric composition a more tractable diagnostic of interior properties compared to planets with smaller cores.

    \item The phase-curve observations of Program 8864 provide a natural framework for testing these predictions and progressively constraining the interior and atmospheric states of each target. Future observations will therefore test the model predictions and help identify where the adopted physical framework may require refinement.

\end{itemize}

Taken together, these results demonstrate that comparative phase curve observations of rocky exoplanets, even in the absence of spectral features, carry significant diagnostic power for constraining interior and atmospheric states. The diversity of predicted outcomes across the five targets of Programme 8864  spanning fully molten, partially molten, and solid interiors, with and without atmospheres  underscores the value of a multi-planet survey approach for disentangling the parameter degeneracies inherent to single-planet studies. As JWST observations of these targets become available, the framework presented here provides a direct basis for interpreting the data and advancing our understanding of the boundary conditions that govern atmospheric retention and 
interior evolution on rocky exoplanets.

\begin{acknowledgements}
This research was supported by the Branco Weiss Foundation, the European Research Council (ERC) under the European Union’s Horizon Europe research and innovation programme (MagmaWorlds, 101219807), the Alfred P. Sloan Foundation (AEThER, G-2025-25284), NASA’s Nexus for Exoplanet System Science research coordination network (Alien Earths, 80NSSC21K0593), and the NWO NWA-ORC PRELIFE Consortium (NWA.1630.23.013). MS thanks Quentin Changeat for useful comments and suggestions. J.P.W. acknowledges support from the Swiss National Science Foundation (SNSF) under grant 10002706 (PIs: D. Kitzmann, H.J. Hoeijmakers) and from the Canadian Space Agency (CSA) under grant 24JWGO3A-03. Y.M. acknowledges funding from the European Research Council (ERC) under the European Union’s Horizon 2020 research and innovation programme (grant agreement no. 101088557, N-GINE). O.H. acknowledges financial support by FFG under grant number 62545255. M.L-M. is supported by individual research time under NASA contracts NAS5-26555 and NAS5-03127 to the Associated Universities for Research in Astronomy for the operation of the Hubble Space Telescope and James Webb Space Telescope Science Operations Centers at STScI.
M.H would like to thank the Fonds de recherche du Qu\'ebec for a doctoral fellowship. H.S.W acknowledges support from the Carlsberg Foundation through the FIRSTATMO project. S.Z. was supported by NASA through the NASA Hubble Fellowship grant \#HST-HF2-51570.001-A awarded by the Space Telescope Science Institute, which is operated by the Association of Universities for Research in Astronomy, Incorporated, under NASA contract NAS5-26555. A.A.A.P. acknowledges funding from a UK Science and Technology Facilities Council (STFC) Small Award, grant number UKRI/ST/B001171/1. We also thank the Center for Information Technology of the University of Groningen for their support and for providing access to the Habrok high performance computing cluster.

\end{acknowledgements}

\bibliography{references}{}
\bibliographystyle{aasjournalv7}

\end{document}